\documentclass[twocolumn,11pt]{aa}
\PassOptionsToPackage{usenames,dvipsnames,svgnames,table}{xcolor}
\pdfoutput=1
\usepackage[varg]{txfonts}

\usepackage{xcolor}

\usepackage{graphicx}
\usepackage{xspace}
\usepackage{amsmath}
\usepackage{mathtools}
\usepackage{enumitem}

\usepackage[colorlinks=true, linkcolor=blue, citecolor=blue, filecolor=blue, urlcolor=blue]{hyperref}

\newcommand{\unit}[1]{\mathrm{\,#1}}
\newcommand{\Msun}{\ensuremath{M_{\odot}}\xspace}

\newcommand{\Mstar}{\ensuremath{M_*}\xspace}
\newcommand{\Mgas}{\ensuremath{M_{\mathrm{gas}}}\xspace}
\newcommand{\Mbar}{\ensuremath{M_{\mathrm{bar}}}\xspace}
\newcommand{\Mhalo}{\ensuremath{M_{\mathrm{halo}}}\xspace}

\newcommand{\lMstar}{\ensuremath{\log_{10}(\Mstar/\Msun)}\xspace}

\newcommand{\lMhalo}{\ensuremath{\log_{10}(\Mhalo/\Msun)}\xspace}
\newcommand{\chalo}{\ensuremath{c_{\mathrm{halo}}}\xspace}

\newcommand{\vrot}{\ensuremath{v_{\mathrm{rot}}}\xspace}

\newcommand{\vcirc}{\ensuremath{v_{\mathrm{circ}}}\xspace}
\newcommand{\vcirctot}{\ensuremath{v_{\mathrm{circ,tot}}}\xspace}
\newcommand{\vcircDM}{\ensuremath{v_{\mathrm{circ,DM}}}\xspace}

\newcommand{\sigmaint}{\ensuremath{\sigma_{0}}\xspace}

\newcommand{\bt}{\ensuremath{B/T}\xspace}

\newcommand{\fgas}{\ensuremath{f_{\mathrm{gas}}}\xspace}  

\newcommand{\fDM}{\ensuremath{f_{\mathrm{DM}}}\xspace}  
\newcommand{\fDMv}{\ensuremath{f_{\mathrm{DM}}^{v}}\xspace}  
\newcommand{\fDMm}{\ensuremath{f_{\mathrm{DM}}^{m}}\xspace}

\newcommand{\Reff}{\ensuremath{R_{\mathrm{e}}}\xspace}  
\newcommand{\Redisk}{\ensuremath{R_{\mathrm{e,disk}}}\xspace}
\newcommand{\Rebulge}{\ensuremath{R_{\mathrm{e,bulge}}}\xspace}

\newcommand{\nSdisk}{\ensuremath{n_{\mathrm{disk}}}\xspace}  
\newcommand{\nSbulge}{\ensuremath{n_{\mathrm{bulge}}}\xspace}  

\newcommand{\rhalftD}{\ensuremath{r_{1/2,\mathrm{mass,3D}}}\xspace}
\newcommand{\rhalftDbar}{\ensuremath{r_{1/2,\mathrm{3D,baryons}}}\xspace}
\newcommand{\bn}{\ensuremath{b_{\mathrm{n}}}\xspace}

\newcommand{\ktot}{\ensuremath{k_{\mathrm{tot}}}\xspace}
\newcommand{\kthrd}{\ensuremath{k_{\mathrm{3D}}}\xspace}

\newcommand{\qint}{\ensuremath{q_0}\xspace}
\newcommand{\qintdisk}{\ensuremath{q_{0,\mathrm{disk}}}\xspace}
\newcommand{\qintbulge}{\ensuremath{q_{0,\mathrm{bulge}}}\xspace}
\newcommand{\qobs}{\ensuremath{q_{\mathrm{obs}}}\xspace}

\newcommand{\Mtot}{\ensuremath{M_{\mathrm{tot}}}\xspace}
\newcommand{\Ltot}{\ensuremath{L_{\mathrm{tot}}}\xspace}

\newcommand{\Menc}{\ensuremath{M_{\mathrm{sph}}}\xspace}
\newcommand{\MencDM}{\ensuremath{M_{\mathrm{DM,sph}}}\xspace}
\newcommand{\Menctot}{\ensuremath{M_{\mathrm{tot,sph}}}\xspace}
\newcommand{\Mencspheroid}{\ensuremath{M_{\mathrm{spheroid}}}\xspace}

\newcommand{\alphaSG}{\ensuremath{\alpha_{\mathrm{self-grav}}}\xspace}
\newcommand{\alphan}{\ensuremath{\alpha(n)}\xspace}
\newcommand{\alphatot}{\ensuremath{\alpha_{\mathrm{tot}}}\xspace}

\newcommand{\D}{\ensuremath{\mathrm{d}}\xspace}

\defcitealias{Noordermeer08}{N08}

\newcommand*{\MPE}{Max-Planck-Institut f\"{u}r extraterrestrische Physik (MPE), Giessenbachstr. 1, D-85748 Garching, Germany}
\newcommand*{\UCB}{Departments of Physics and Astronomy, University of California, Berkeley, CA 94720, USA}
\newcommand*{\Cavendish}{Cavendish Laboratory, University of Cambridge, 19 J.J. Thomson Avenue, Cambridge CB3 0HE, UK}
\newcommand*{\KICC}{Kavli Institute for Cosmology, University of Cambridge, Madingley Road, Cambridge CB3 0HA, UK}
\newcommand*{\UAF}{Physics Department, University of Alaska, Fairbanks, AK 99775, USA}
\newcommand*{\USM}{Universit\"{a}ts-Sternwarte Ludwig-Maximilians-Universit\"{a}t (USM), Scheinerstr. 1, D-81679 M\"{u}nchen, Germany}
\newcommand*{\Leiden}{Sterrewacht Leiden, Leiden University, Postbus 9513, 2300 RA Leiden, The Netherlands}
\newcommand*{\UWC}{University of the Western Cape, Bellville, Cape Town 7535, South Africa}

\titlerunning{Kinematics and Mass Distributions for Non-Spherical Deprojected S\'ersic Density Profiles}
\authorrunning{Price et al.}

\begin{document}

\title{Kinematics and Mass Distributions for\\Non-Spherical Deprojected S\'ersic Density Profiles\\and Applications to Multi-Component Galactic Systems}

\author{S.~H.~Price\inst{\ref{MPE}} \and 
H.~\"{U}bler\inst{\ref{Cavendish}, \ref{KICC}} \and 
N.~M.~F\"{o}rster~Schreiber\inst{\ref{MPE}} \and
P.~T.~de~Zeeuw\inst{\ref{MPE}, \ref{Leiden}} \and
A.~Burkert\inst{\ref{MPE}, \ref{USM}} \and \\
R.~Genzel\inst{\ref{MPE}, \ref{UCB}} \and 
L.~J.~Tacconi\inst{\ref{MPE}} \and 
R.~I.~Davies\inst{\ref{MPE}} \and 
C.~P.~Price\inst{\ref{UAF}, \ref{UWC}}}

\institute{\MPE\\\email{sedona@mpe.mpg.de}\label{MPE} 
\and \Cavendish\label{Cavendish} 
\and \KICC\label{KICC} 
\and \Leiden\label{Leiden} 
\and \USM\label{USM} 
\and \UCB\label{UCB} 
\and \UAF\label{UAF} \and \UWC\label{UWC} 
}

\date{\today; Accepted to \emph{Astronomy \& Astrophysics}}


\abstract{
\noindent
Using kinematics to decompose galaxies' mass profiles, including the dark matter contribution, 
often requires parameterization of the baryonic mass distribution based on ancillary information. 
One such model choice is a deprojected S\'ersic profile with an assumed intrinsic geometry.
The case of flattened, deprojected S\'ersic models has previously been applied to flattened bulges 
in local star-forming galaxies (SFGs), but can also be used to describe the thick, turbulent disks in distant SFGs. 
Here we extend this previous work that derived density ($\rho$) and circular velocity (\vcirc) curves
by additionally calculating the spherically-enclosed 3D mass profiles (\Menc). 
Using these profiles, we compare the projected and 3D mass distributions,  
quantify the differences between the projected and 3D half-mass radii (\Reff; \rhalftD), 
and present virial coefficients relating $\vcirc(R)$ and $\Menc(<r=R)$ or \Mtot. 
We then quantify differences between mass fraction estimators 
for multi-component systems, particularly for dark matter fractions 
(ratio of squared circular velocities versus ratio of spherically enclosed masses), 
and consider the compound effects of measuring dark matter fractions 
at the projected versus 3D half-mass radii. 
While the fraction estimators produce only minor differences, 
using different aperture radius definitions can strongly impact the inferred dark matter fraction.
As pressure support is important in analysis of gas kinematics (particularly at high redshifts), 
we also calculate the self-consistent pressure support correction profiles, 
which generally predict less pressure support than for the self-gravitating disk case. 
These results have implications for comparisons between simulation and observational measurements, 
and for the interpretation of SFG kinematics at high redshifts. 
A set of precomputed tables and the code to calculate the profiles are made publicly available.}


\maketitle

\begin{table*}[ht]
\small
\caption{Definitions of key variables}
\label{tab:variables}
\begin{centering}
\vspace{-9pt}
\begin{tabular}{c l l }
\hline\hline
Variable & Definition & Reference  \\
\hline
 & \emph{--- Model ---}  &  \\
$n$ & S\'ersic index & Sec.~\ref{sec:rhom} \\
$\Reff$ & Projected 2D S\'ersic effective radius & Sec.~\ref{sec:rhom}  \\
$\qint$ & Intrinsic axis ratio $c/a$ & Sec.~\ref{sec:rhom}  \\[1pt]
\arrayrulecolor{gray}\hline
&  \emph{--- Geometry ---}   &  \\
$R$ & Radius in the midplane & Sec.~\ref{sec:rhom}  \\
$z$ & Height above the midplane & Sec.~\ref{sec:rhom}  \\
$m = \sqrt{R^2 + (z/\qint)^2}$ & Spheroid isodensity surface distance & Sec.~\ref{sec:rhom} \\[1pt]
\hline 
&  \emph{--- Derived ---}    &  \\
$\rho(m)$ & 3D deprojected density & Sec.~\ref{sec:rhom}  \\
$\vcirc(R)$ & Circular velocity in the midplane, accounting for non-spherical potentials & Sec.~\ref{sec:vc} \\
$\Menc(<r=R)$ & Mass enclosed within a sphere of radius $r=R$ & Sec.~\ref{sec:mencl}  \\
$\Mencspheroid(<m=R)$ & Mass enclosed within spheroid with isodensity surface distance $m=R$ 
and intrinsic axis ratio $\qint$ & Sec.~\ref{sec:mencl} \\
\rhalftD & 3D spherical half-mass radius (assuming constant $M/L$) & Sec.~\ref{sec:menclvcirc_props} \\
$\kthrd(\Reff)$ & 3D enclosed mass virial coefficient relating $\vcirc(R)$ to $\Menc(<r=R)$ & Sec.~\ref{sec:virialcoeff} \\
$\ktot(R)$ & Total virial coefficient relating $\vcirc(R)$ to \Mtot & Sec.~\ref{sec:virialcoeff} \\[1pt]
$\fDMm(R)$ & Dark matter fraction defined as ratio of dark matter to total mass enclosed within a sphere of radius $r=R$ & Sec.~\ref{sec:deffDM} \\[1pt]
$\fDMv(R)$ & Dark matter fraction defined as ratio of dark matter to total circular velocities squared at radius $R$ & Sec.~\ref{sec:deffDM} \\
$\alpha(R)$ & Pressure support correction ($=\D\ln\rho_g/\D\ln{}R$ for constant dispersion) & Sec.~\ref{sec:asymmdrift_single} \\[1pt]
\arrayrulecolor{black}\hline
\end{tabular}
\end{centering}
\vspace{-9pt}
\end{table*}


\section{Introduction}
\label{sec:intro}

Galaxy kinematics, such as rotation curves, are a powerful tool to measure the mass of all components in a galaxy 
(e.g., \citealt{vanderKruit78}, \citealt{Courteau14}). 
Notably,  this technique has been used to study 
the dark matter content of galaxies at a wide range of epochs, including constraints on the halo profile shapes 
(e.g., \citealt{Sofue01}, \citealt{deBlok10}, \citealt{Genzel20}, among many others). 
Furthermore, by using kinematics to probe the mass and angular momentum distribution 
within galaxies, it is possible to constrain the processes regulating galaxy growth 
and evolution over time (\citealt{vanderKruit11}, \citealt{ForsterSchreiber20}; 
see also, e.g., \citealt{Mo98}, \citealt{Sofue01}, \citealt{Romanowsky12}). 
It is especially informative to study the kinematics of star-forming galaxies (SFGs), which tend to lie 
on a tight ``star-forming main sequence'' where much of cosmic star formation occurs 
(\citealt{Speagle14};  \citealt{Rodighiero11}, \citealt{Whitaker14}, \citealt{Tomczak16}).
However, there are challenges to recovering the intrinsic mass properties of galaxies from their observed kinematics.

One such challenge is that in order to overcome degeneracies in kinematic mass decomposition 
(particularly when including an unseen dark component; e.g., \citealt{vanAlbada85}), 
separate constraints on the baryonic (gas and stellar) component are needed, 
either through empirical measurements or with a choice of parameterization 
(e.g., \citealt{Persic96}, \citealt{deBlok97},  \citealt{Palunas00}, \citealt{Dutton05}, \citealt{deBlok08}, \citealt{Courteau14}). 
Multi-wavelength imaging and spectroscopy (in emission or absorption) can constrain the distribution of gas and stars in galaxies. 
Such observations of individual galaxies provide projected information and not the 3D quantities needed for kinematic modeling. 
Consequently, it is often necessary to first parameterize the projected distributions 
and then make reasonable assumptions about the galaxies' intrinsic geometries 
in order to deproject the surface distributions into 3D mass profiles.

Observationally, the light distributions of galaxies are often described by \citet{Sersic68} profiles 
(e.g., \citealt{Peng02,Peng10a}, \citealt{Simard02,Simard11}, \citealt{Blanton03}, \citealt{Wuyts11b}, 
\citealt{vanderWel12}, \citealt{Conselice13}, and numerous others). 
In some cases, there are distinct components within galaxies, but these are also frequently described 
by S\'ersic profiles with distinct indices $n$ and effective radii \Reff (e.g., a disk and bulge for star-forming galaxies; 
\citealt{Courteau96b}, \citealt{Bruce12}, \citealt{Lang14}). 
Thus, S\'ersic profiles are a natural choice for the projected parameterization.

Deprojections of S\'ersic profiles have been studied in numerous previous works, 
for spherical (e.g., \citealt{Ciotti91}, \citealt{Ciotti97}, \citealt{Baes19, Baes19b}), 
triaxial (e.g., \citealt{Stark77}, \citealt{Trujillo02}), 
and axisymmetric geometries (e.g., \citealt{Noordermeer08}). 
Additionally, the dynamics for exponential surface profiles 
have been derived for both razor-thin (\citealt{Freeman70}) and finitely thick (\citealt{Casertano83}) geometries
(though these are generalizable to arbitrary S\'ersic index; e.g., see \citealt{BT08}). 
These intrinsic geometries have applications for various galaxies or galaxy components, 
depending on the galaxy properties and epoch.

In particular, the mass distribution geometry of SFGs changes over time. 
Nearby SFGs often have thin disks, particularly in the gas components (\citealt{vanderKruit11}), 
while distant (massive) SFGs tend to have thick, turbulent disks 
(\citealt{Glazebrook13}, \citealt{ForsterSchreiber20}, and references therein). 
While more observations are needed to better constrain the vertical disk structure of distant, massive SFGs, 
flattened (oblate) distributions are more appropriate models 
(as adopted by, e.g., \citealt{Wuyts16}, \citealt{Genzel17, Genzel20}), 
using the same geometric deprojection used by \citet{Noordermeer08} to describe the flattening of nearby bulges.

A second challenge is that the observed rotation must be corrected for pressure support. 
This correction is important for gas kinematic measurements, especially at high redshifts where disks have high gas turbulence.
A number of works have considered different analytic prescriptions for correcting for the pressure support 
in gas kinematics (e.g., \citealt{Weijmans08}, 
\citealt{Burkert10}, \citealt{Dalcanton10}, \citealt{Kretschmer21}). 
In general, such corrections require measurements of the gas turbulence $\sigma$ 
from spatially-resolved spectroscopy (i.e., slit along the major axis or kinematic maps) 
as well as constraints or parameterizations of the gas density profile. 
If deprojected S\'ersic distributions are used to model the 
mass and \vcirc profiles for galaxies' gas and stellar components, 
then a pressure support prescription derived using 
the density slope can be adopted for a self-consistent kinematic analysis 
(as in, e.g., \citealt{Weijmans08}, \citealt{Burkert10}, \citealt{Dalcanton10}). 
If galaxies exhibit non-constant dispersion, support from dispersion gradients or anisotropy 
can also be included (e.g., \citealt{Weijmans08}, \citealt{Dalcanton10}).

In order to further consider implications for the interpretation of the kinematics of high-redshift SFGs 
modeled using deprojected, flattened S\'ersic profiles, in this paper we revisit and extend the framework 
first presented by \citet[hereafter \citetalias{Noordermeer08}]{Noordermeer08}. 
We first present various profile derivations for deprojected, flattened S\'ersic profiles, 
including the density and circular velocity profiles determined by \citetalias{Noordermeer08} 
as well as the spherically-enclosed 3D mass profiles (Sec.~\ref{sec:deriv}). 
Using the calculated profiles, we examine the relationship between projected 2D and 3D mass distributions, 
including differences between the 2D \Reff and 3D \rhalftD  (Sec.~\ref{sec:menclvcirc_props}). 
The circular velocity and 3D mass distributions are also used to calculate virial coefficients (Sec.~\ref{sec:virialcoeff}). 
Next, we examine the circular velocity and enclosed mass profiles for multi-component systems 
for a range of realistic $z\sim2$ galaxy properties (Sec.~\ref{sec:mdistgal}).
We find the composite baryonic 3D half-mass radius \rhalftDbar is often smaller than the projected disk effective radius \Redisk. 
While different dark matter fraction estimators 
\fDMv{} (the ratio of the dark matter to total circular velocities squared) 
and \fDMm{} (the ratio of the dark matter to total mass enclosed within a sphere) 
are similar when calculated at the same radius, 
large differences in \fDM can result from the use of different aperture radii (\rhalftDbar vs. \Redisk). 
We then determine the self-consistent turbulent pressure support correction, assuming a constant \sigmaint, 
which is typically only half the amount predicted for a self-gravitating disk, 
and demonstrate the correction for a range of realistic $z\sim2$ galaxy properties (Sec.~\ref{sec:asymmdrift}).  
Finally, we discuss these results and their implications, in particular for comparisons between simulations and observations
and for studies of disk galaxy kinematics at $z\sim1-3$ (Sec.~\ref{sec:disc}). 
We highlight how typical observational and simulation ``half-mass'' radius estimates 
can lead to differences of up to $\sim0.15$ in measured \fDM, and how 
the lower pressure support correction derived for these mass distributions (compared to the self-gravitating disk prescription) 
would imply typically lower inferred \fDM from observations.

\begin{figure*}[t!] 
\vspace{-6pt}
\centering
\hglue -4pt
\includegraphics[width=1.015\textwidth]{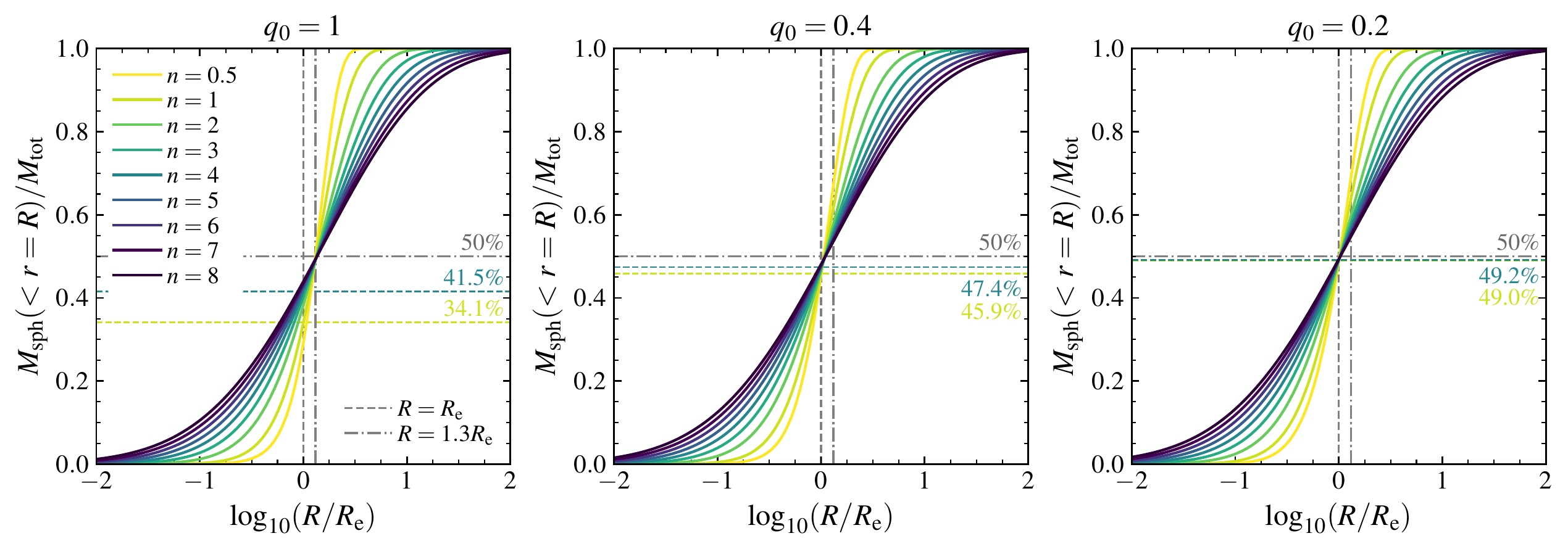} 
\vspace{-18pt}
\caption{Fractional mass enclosed within a sphere of radius $r=R$ for 
deprojected S\'ersic models of different intrinsic axis ratios. 
From left to right, the enclosed \Menc is plotted as a function of log radius 
(relative to the projected 2D effective radius, \Reff), assuming intrinsic axis ratios of $\qint =1, 0.4, 0.2$, respectively. 
The colored curves denote the enclosed mass profiles for S\'ersic indices from $n=0.5$ to $n=8$ (yellow to purple).
The vertical lines denote $R=\Reff$ (grey dashed) and $R=1.3\Reff$ ($\approx \rhalftD$ for $\qint=1$; grey dashed dotted), 
and the horizontal colored lines denote the fraction of the mass enclosed within $r=\Reff$ for $n=1, 4$ 
(lime, teal dashed, respectively) and 50\% of the total mass (grey dashed dotted). 
For $\qint=1$, the half-mass 3D spherical radius is indeed $\rhalftD\approx1.3\Reff$ regardless of $n$, as in \citet{Ciotti91}. 
For flattened (i.e., oblate) systems, the half-mass 3D spherical radius is smaller, 
and approaches $\Reff$ as $\qint$ decreases. See also Fig.~\ref{fig:2}.}
\vspace{-7pt}
\label{fig:1}
\end{figure*}

A set of tables containing precomputed profiles and values  --- including 
$\vcirc(R)$ (Eq.~\ref{eq5}), $\Menc(\mbox{$<r=R$})$ (Eq.~\ref{eq8}), 
$\Mencspheroid(\mbox{$<m=R$})$ (Eq.~\ref{eq6}), $\rho(r=R)$ (Eq.~\ref{eq2}), 
$\D\ln\rho/\D\ln{}R$ (derived from Eq.~\ref{eq17}), $\rhalftD$, 
$\ktot(\Redisk)$ (Eq.~\ref{eq10}), and $\kthrd(\Redisk)$ (Eq.~\ref{eq9}) --- 
for a range of intrinsic axis ratios \qint and S\'ersic indices, 
and the code used to compute the profiles, are made available.\footnote{The \texttt{python} package 
\texttt{deprojected\_sersic\_models} used in this paper 
and the pre-computed tables are both available for download from 
\href{https://sedonaprice.github.io/deprojected_sersic_models/downloads.html}{sedonaprice.github.io/deprojected\_sersic\_models/downloads.html}; 
the full code repository is publicly available at 
\href{https://github.com/sedonaprice/deprojected_sersic_models/releases}{github.com/sedonaprice/deprojected\_sersic\_models}. 
The code also includes 
functions for 
scaling and interpolating the profiles from the pre-computed tables to arbitrary 
total masses and \Reff as a function of radius.} 
For reference, key variables and their definitions are listed in Table~\ref{tab:variables}.

We assume a flat $\Lambda$CDM cosmology with $\Omega_m = 0.3$, 
$\Omega_{\Lambda} = 0.7$, and $H_0 = 70\unit{km\ s^{-1}\ Mpc^{-1}}$.


\vspace{-2pt} 

\section{Derivation of mass profiles and rotation curves}
\label{sec:deriv}

In this section, we present formulae for the mass profiles and rotation curves 
for models whose projected intensity distributions follow S\'ersic profiles, 
but that have oblate (flattened) or prolate axisymmetric 3D density distributions 
(i.e., the isodensity contours follow oblate/prolate spheroids), 
following the deprojection derivation of \citetalias{Noordermeer08}.

\vspace{-3pt}

\subsection{Deprojected S\'ersic density profile}
\label{sec:rhom}

We assume that the mass density of the 3D spheroid can be written as $\rho(x,y,z) = \rho(m)$,  
where $(m/a)^2 = (x/b)^2 + (y/a)^2 + (z/c)^2$ specifies the isodensity surfaces 
for a given set of semi-axis lengths $a, b, c$. 
For an axisymmetric system this simplifies to $m=\sqrt{R^2+(z/\qint)^2}$, where $R=\sqrt{x^2+y^2}$ 
is the distance in the plane of axisymmetry,  $z$ is the distance from the midplane, 
the semi axes $a=b$, and $\qint=c/a$ is the intrinsic axis ratio of the spheroid. 
To project the intrinsic galaxy coordinates to the observer's frame, we adopt the transformation from 
$(x,y,z)$ to $(\zeta, \kappa, \xi)$ from Eq.~1 of \citetalias{Noordermeer08}, where $\zeta$ lies along the line-of-sight, 
$\kappa$ and $\xi$ lie along the galaxy major and minor axes (as viewed in the sky plane, 
for oblate geometries\footnote{For prolate geometries, 
the projected major axis lies parallel to the long intrinsic axis, $c$. 
Here, however, we use a geometry definition where $\kappa$ is parallel to $a$ for all cases, 
for a consistent convention relative to the rotation axis ($z$; parallel to $c$) --- so technically $\kappa$ 
is parallel to the major axis as usual for oblate geometry, but lies along the minor axis for prolate geometry.}; i.e., $\kappa=a$), 
respectively, and $i$ is the inclination of the system relative to the observer (see also Fig.~1 of \citetalias{Noordermeer08}). 
The observed axis ratio of the ellipsoid is then 
$\qobs = \sqrt{\qint^2 + (1-\qint^2)\cos^2i}$.

Within the observer's coordinate frame, the relationship between the 3D mass density profile 
and the projected light intensity 
along the major axis of the galaxy is ($\xi=0$; from Eq.~8 of \citetalias{Noordermeer08}\footnote{In 
\citetalias{Noordermeer08}, $\rho(m)$ denotes the 3D luminosity density distribution, 
while we define $\rho(m)$ as the 3D mass density. 
Thus, we instead write the 3D luminosity density as $\rho(m)/\Upsilon(m)$ 
in the projection integral.}): 
\begin{equation}
I(\kappa)=2\frac{\qint}{\qobs}\int_{\kappa}^{\infty}\frac{1}{\Upsilon(m)}\rho(m)\frac{m\,\D{}m}{\sqrt{m^2-\kappa^2}},
\label{eq1}
\end{equation} 
where $\Upsilon(m)$ is the mass-to-light ratio of the galaxy and $\qobs/\qint = \sqrt{\sin^2i + (1/\qint^2) \cos^2i}$. 
For simplicity, we assume a constant mass-to-light ratio, $\Upsilon(m) \equiv \Upsilon$.

\begin{figure*}[t!]
\vspace{-6pt}
\centering
\includegraphics[width=0.995\textwidth]{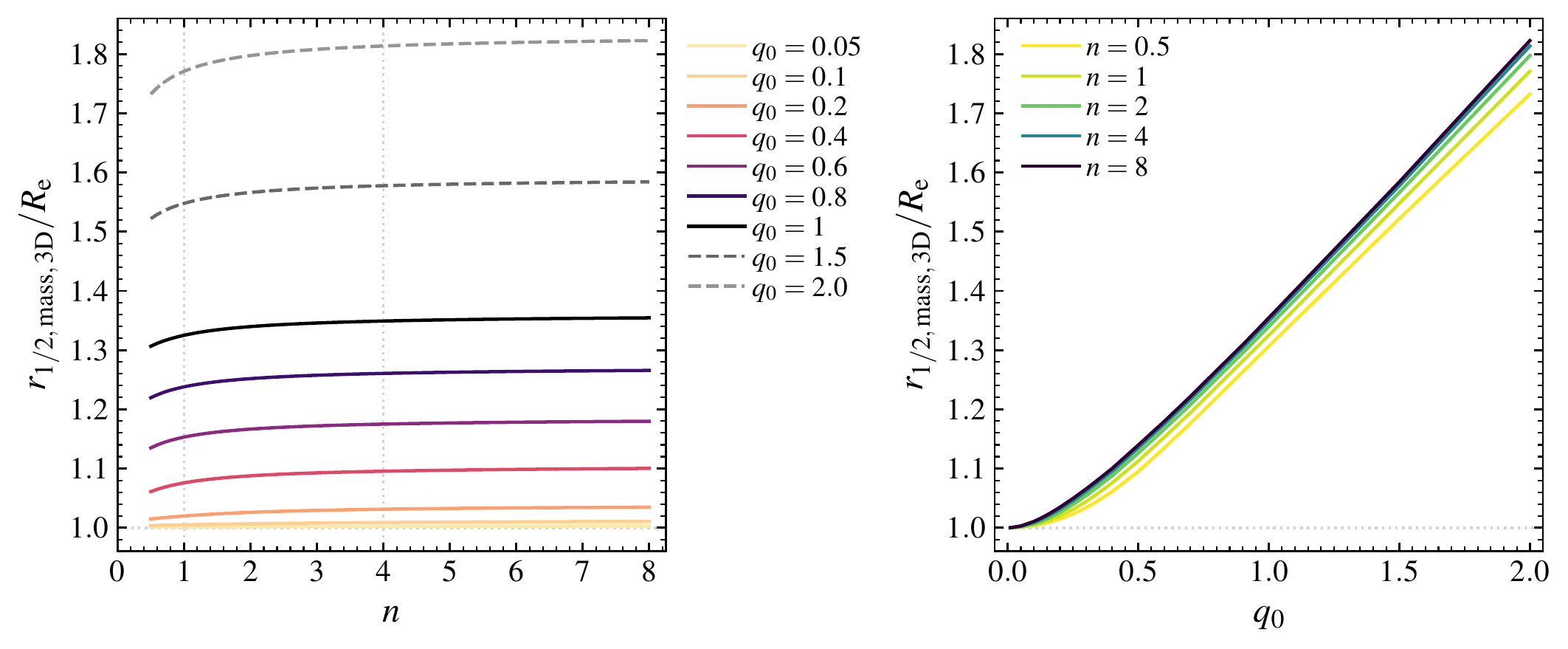}
\vspace{-10pt}
\caption{Comparison between the 3D spherical half-mass radius, $\rhalftD$ and the 
projected 2D effective radius, $\Reff$, for a range of S\'ersic indices $n$ and intrinsic axis ratios $\qint$ 
(\emph{left}, colored by $\qint$; \emph{right}, colored by $n$). 
For oblate cases, $\Reff$ is the projected major axis, while for prolate cases $\Reff$ is the projected minor axis. 
For all cases, $\rhalftD > \Reff$. 
However, as $\qint$ decreases (i.e., flatter S\'ersic distributions), the 3D half-mass radius approaches the value of \Reff. 
Overall, the systematic difference between $\rhalftD$ and $\Reff$ emphasizes that, while half of the model 
mass is enclosed within a \emph{projected 2D ellipse} of major axis \Reff (e.g., an infinite ellipsoidal cylinder), 
less than half the total mass is enclosed within a \emph{sphere} of radius \Reff (ignoring any $M/L$ gradients or 
optically thick regions, which would change $R_{e,\mathrm{light}}/R_{e,\mathrm{mass}}$).}
\label{fig:2}
\vspace{-7pt}
\end{figure*}

The deprojected density profile is found by inverting this Abel integral, with (c.f. Eq.~9, \citetalias{Noordermeer08}) 
\begin{equation}
\rho(m)=-\frac{\Upsilon}{\pi}\frac{\qobs}{\qint}\int_{m}^{\infty} \frac{\D{}I}{\D{}\kappa}\frac{\D{}\kappa}{\sqrt{\kappa^2-m^2}}.
\label{eq2}
\end{equation} 
We write the S\'ersic profile as (c.f. Eq.~11, \citetalias{Noordermeer08}) 
\begin{equation}
I(\kappa)=I_e\exp\left\{-\bn\left[\left(\frac{\kappa}{\Reff}\right)^{1/n}-1\right]\right\},
\label{eq3}
\end{equation} 
where $\Reff$ is the effective radius, $n$ is the S\'ersic index, $I_e$ is the surface brightness at $\Reff$,  
and $\bn$ satisfies 
$\gamma(2n,\bn)=\frac{1}{2}\,\Gamma(2n)$, 
\parbox{0.49\textwidth}{\sloppy\setlength\parfillskip{0pt}where $\Gamma(a)$ and $\gamma(a, x)$ are the regular and 
lower incomplete gamma functions, respectively (e.g., \citealt{Graham05a}).}

\par \noindent
The derivative is then 
\begin{equation}
\frac{\D{}I}{\D{}\kappa}=-\frac{I_e\bn}{n\Reff}\exp\left\{-\bn\left[\left(\frac{\kappa}{\Reff}\right)^{1/n}-1\right]\right\}
\left(\frac{\kappa}{\Reff}\right)^{(1/n)-1}.
\label{eq4}
\end{equation} 
By inserting Eq.~\ref{eq4} into Eq.~\ref{eq2}, we can numerically integrate to obtain the deprojected density profile $\rho(m)$.  
For the adopted convention here, where the projected $\kappa$ lies along $a$ in the midplane
(so this is the usual projected major axis for oblate cases but is the projected \emph{minor} axis for prolate cases), 
we have $\Reff=a$ as the projected effective radius.

\subsection{Rotation curves}
\label{sec:vc}

Next we determine the circular rotation curve for this class of density profiles, following the derivation of 
\citet[Eq.~2.132; also Eq.~10, \citetalias{Noordermeer08}]{BT08}. 
The circular rotation curve at the midplane of the galaxy is thus 
\begin{align}
\begin{split}
\vcirc^2(R)=-&4G\qobs\Upsilon\\
\times\int_{m=0}^{R}&\left[\int_{\kappa=m}^{\infty}\frac{\D{}I}{\D{}\kappa}\frac{\D{}\kappa}{\sqrt{\kappa^2-m^2}}\right]\frac{m^2\,\D{}m}{\sqrt{R^2-(1-\qint^2)m^2}} .
\label{eq5}
\end{split}
\end{align} 
As noted by \citetalias{Noordermeer08}, this equation is valid for any observed intensity profile $I(\kappa)$. 
Here we combine Eqs.~\ref{eq2}~and~\ref{eq5}, which can be numerically integrated to yield $\vcirc(R)$.

\subsection{Enclosed 3D mass}
\label{sec:mencl}

We next derive the enclosed mass for models with the density profiles given above. 
Given the modified coordinate $m$, the mass enclosed within a spheroid 
with intrinsic axis ratio \qint can be expressed as 
\begin{align}\begin{split} 
\Mencspheroid&(<m=R)\equiv M_{\mathrm{3D,spheroid}}(<m=R)\\
&=4\pi\qint\int_{0}^{R}m^2\,\D{}m\,\rho(m)\,\,. 
\label{eq6}
\end{split}
\end{align} 
Integrated to infinity, this is equivalent to the total luminosity of the S\'ersic profile 
times the constant assumed mass-to-light ratio, or $\Mtot = \Upsilon \Ltot$, 
so the intensity normalization for a flattened S\'ersic profile with observed axis ratio \qobs is
\begin{equation}I_e = \frac{\Mtot}{\Upsilon{}}\,\frac{1}{\qobs}\,\frac{\bn^{2n}}{2\pi{}\Reff^2n\,e^{\bn}\,\Gamma(2n)}\,\,.\label{eq7}\end{equation}

However, there may be situations where 
we wish to compute the mass enclosed within a sphere of radius $r=\sqrt{R^2+z^2}$ 
instead of within a flattened (or prolate) spheroid. 
We thus use a change of coordinates to calculate the spherical enclosed mass: 
\begin{align}\begin{split} 
\Menc&(<r=R)\equiv M_{\mathrm{3D,sphere}}(<r=R) \\
&=4\pi\int_{\tilde{R}\,=\,0}^{R}\tilde{R}\;\D{}\tilde{R}
\int_{z\,=\,0}^{\qint\sqrt{R^2\,-\,\tilde{R}\,^2}}\rho\left(\!\!\sqrt{\tilde{R}\,^2+(z/\qint)^2}\right) \D{}z , 
\label{eq8}
\end{split}\end{align} 
using $\rho(m)$ from Eq.~\ref{eq2}, with $m=\sqrt{\tilde{R}\,^2+(z/\qint)^2}$. 
This integral can be numerically evaluated to find the 3D spherical enclosed mass profile 
corresponding to the deprojected, axisymmetric S\'ersic profile. 
Note that when $\qint\neq1$, then $\Menc(<r=R)\neq v^2_{c}(R) R / G$ 
(with $v^2_{c}(R)$ from Eq.~\ref{eq5}), though the enclosed mass and circular velocity can be related 
through the introduction of a non-unity, radially varying virial coefficient (see Section~\ref{sec:virialcoeff}).

Finally, we note that the mass enclosed within an ellipsoidal cylinder of axis ratio $\qobs$ (and infinite length) 
is equivalent to the enclosed luminosity for the 2D projected S\'ersic profile times the mass-to-light ratio, 
$M_{\mathrm{cyl}}(<\kappa=R) =  2 \pi n \Upsilon I_e \Reff^2  \mathrm{e}^{\bn}(\bn)^{-2n} \gamma(2n,x),$ 
with $x=\bn(R/\Reff)^{1/n}$ (e.g., \citealt{Graham05a}).

\begin{figure*}
\vglue -4pt
\centering
\includegraphics[width=0.85\textwidth]{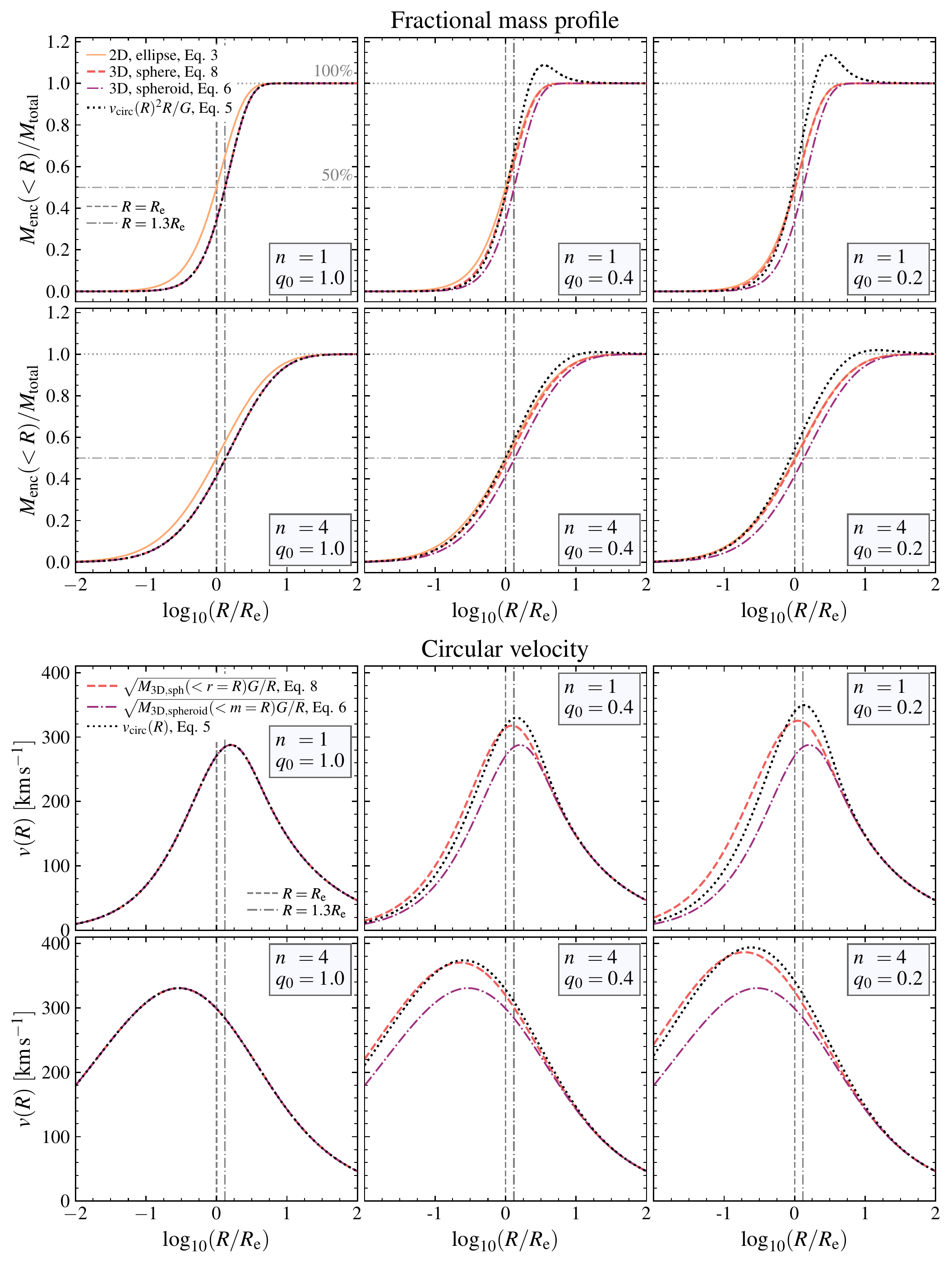}
\vglue -5pt
\caption{Example fractional enclosed mass (\emph{top}) and circular velocity (\emph{bottom}) profiles 
computed or inferred under different assumptions.
The top and bottom rows show the profiles for S\'ersic indices $n=1,4$, respectively, 
while the columns show intrinsic axis ratios $\qint=1 ,0.4, 0.2$ (\emph{from left to right}). 
For the top panels, we show the edge-on 2D projected mass enclosed within ellipses of axis ratio $\qint$ (orange solid line), 
the 3D mass profile enclosed within a sphere (red dashed line), 
the 3D mass profile enclosed within ellipsoids of intrinsic axis ratio $\qint$ (purple dashed dotted line), 
and the mass profile inferred from the flattened deprojected S\'ersic model circular velocity 
under the simplifying assumption of spherical symmetry (i.e., $\qint=1$; black dotted line). 
In the bottom panels, we then compare the flattened deprojected S\'ersic model circular velocity (black dotted line) 
to the inferred velocity profiles computed from the 3D spherical (red solid line) and the 3D ellipsoidal (purple dashed dotted) 
mass profiles under the simple assumption of spherical symmetry. 
The same total mass $\Mtot=5\times10^{10}\Msun$ is used for all cases. 
The vertical lines denote $R=\Reff$ (grey dashed) and $R=1.3\Reff$ ($\approx\rhalftD$ for $\qint=1$; grey dashed dotted). 
These enclosed mass and velocity profiles demonstrate that when $\qint\neq1$, $\Menc(<r=R) \neq \vcirc(R)^2 R / G$. 
The non-spherical potentials for $\qint < 1$ even result in $(\vcirc(R)^2 R / G)/\Mtot > 1$ between $R\sim1-10\Reff$ 
(i.e., potentially leading to $\gtrsim15\%$ overestimates in the system mass). 
We also see that as $\qint$ decreases, \Menc approaches the 2D projected mass profile, 
as the mass enclosed in a sphere versus an infinite ellipsoidal cylinder are equivalent for infinitely thin mass distributions.}
\label{fig:3}
\end{figure*}

\begin{figure*}[ht!] 
\vspace{-6pt}
\centering
\includegraphics[width=0.9\textwidth]{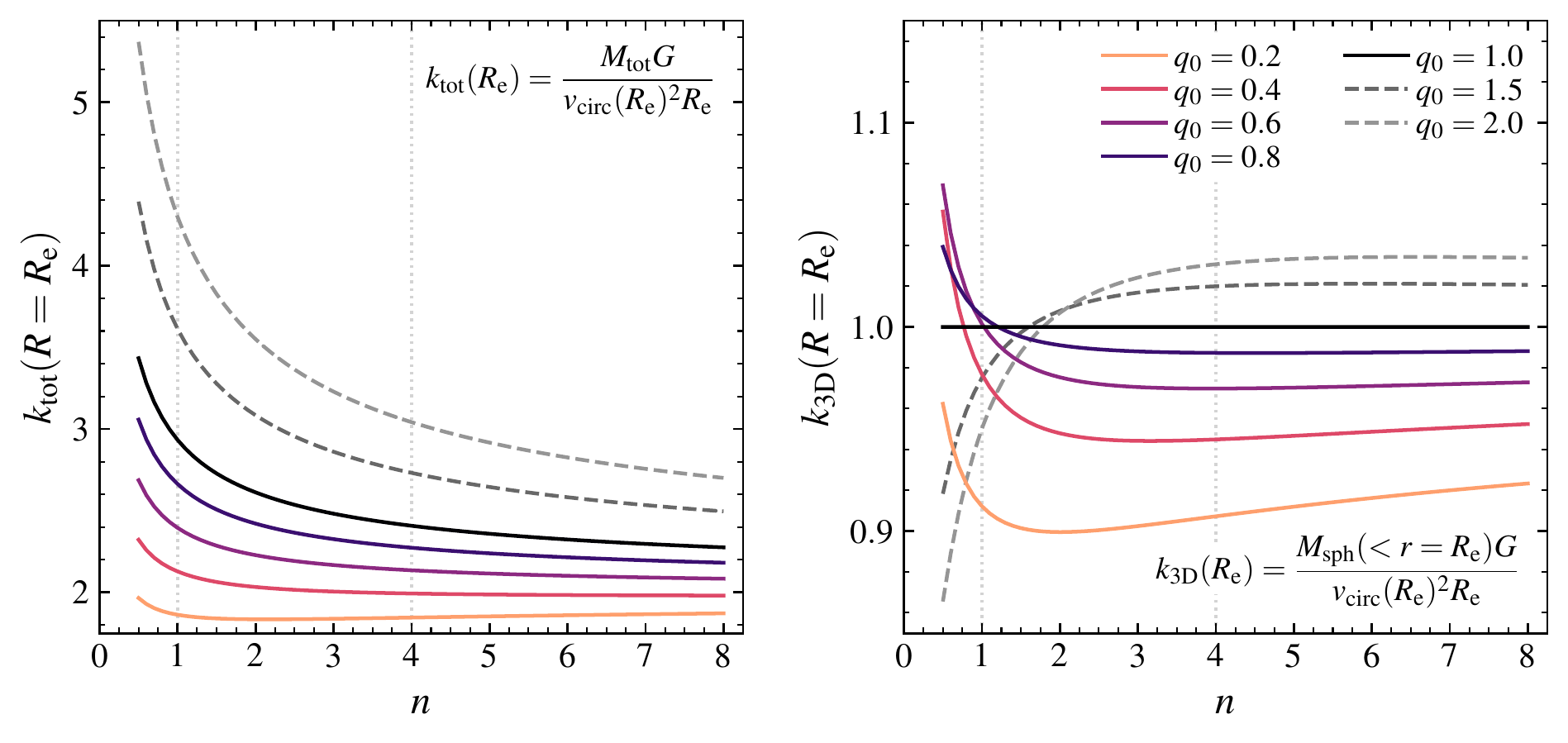}
\vspace{-11pt}
\caption{The total $\ktot(\Reff)$ (\emph{left}) and 3D enclosed $\kthrd(\Reff)$ (\emph{right}) virial coefficients 
as a function of S\'ersic index $n$ and intrinsic axis ratio $\qint$. 
The solid lines denote $\qint=0.2$ (orange) to $\qint=1$ (black), 
and two prolate cases are shown with dashed lines ($\qint=1.5,2$ in dark, light grey, respectively). }
\label{fig:4}
\vspace{-6pt}
\end{figure*}

\begin{table}[t!]
\small
\caption{Virial coefficients for select profiles and radii}
\label{tab:virialcoeff}
\begin{centering}
\vspace{-8pt}
\begin{tabular}{c c || c c | c c }
\hline\hline
 \multicolumn{2}{c}{} & \multicolumn{2}{c}{$\ktot(R)$} & \multicolumn{2}{c}{$\kthrd(R)$} \vspace{2pt}\\
 S\'ersic index & \multicolumn{1}{c}{Axis ratio} & $R=\Reff$ & $1.3\, \Reff$\tablefootmark{a} & 
 $R=\Reff$ & $1.3\, \Reff$\tablefootmark{a}  \\
\hline
$n=1$ & $\qint=0.4$ & 2.128 & 1.512 & 0.977 & 0.929\tablefootmark{b}\\
{       }& $\qint=1\ \ \ $ & 2.933 & 2.026 & 1 & 1 \\
{   $_{\,_{\,_{\,}}}$    }& {\color{Gray}{$\qint=$ \emph{1.5}}} & {\color{Gray}{\emph{3.613}}} &  {\color{Gray}{\emph{2.459}}} &  
 {\color{Gray}{\emph{0.975}}} & {\color{Gray}{\emph{0.995}}}  \\ 
\hline
$n=4$& $\qint=0.4$ & 1.993 &   1.707 & 0.945   &  0.941\tablefootmark{b} \\ 
{       }& $\qint=1\ \ \ $ & 2.408 & 2.033 &    1 & 1 \\ 
{   $_{\,_{\,_{\,}}}$    }& {\color{Gray}{$\qint=$ \emph{1.5}}} & {\color{Gray}{\emph{2.731}}} &  {\color{Gray}{\emph{2.286}}} &   
{\color{Gray}{\emph{1.020}}} & {\color{Gray}{\emph{1.023}}}$_{{}}$ \\
\hline
\end{tabular}\vspace{2pt}
\end{centering}
\tablefoottext{a}{For a $n=1$ S\'ersic profile, $1.3 \, \Reff \approx 2.2 \, R_{s}$.}\\
\tablefoottext{b}{We find $\kthrd(1.3\Reff)=0.746$ and $0.840$ for $n=1, 4$ when 
using the mass enclosed within an ellipsoid instead of a sphere, 
similar to the values $\xi$ for $\qint=0.4$ and 
$n=1,4$ presented by \citet{Miller11} in Sec. 5.1 if their Eq. 6 instead read $M(r) \approx \xi \, \vcirc(r)^2 r /G$.}
\vspace{-8pt}
\end{table}

\subsection{Properties of enclosed mass and circular velocity curves for non-spherical deprojected S\'ersic profiles} 
\label{sec:menclvcirc_props}

The 3D spherical enclosed mass profiles for models with a range of S\'ersic indices ($n=0.5, 1,\ldots,8$) 
and different intrinsic axis ratios $\qint=1, 0.4, 0.2$) are shown in Figure~\ref{fig:1}. 
The 3D spherical half-mass radius (where \rhalftD satisfies $\Menc(<\rhalftD)=\Mtot/2$) 
is $\rhalftD\sim1.3\Reff$ when $\qint=1$ (as shown by \citealt{Ciotti91}).
However, from the $\qint=0.2,0.4$ enclosed mass profiles, we see that the ratio $\rhalftD/\Reff$ 
varies with the model intrinsic axis ratio.

We quantify the dependence of the ratio between the 3D spherical half-mass radius and 
the projected effective radius, $\rhalftD/\Reff$, in Figure~\ref{fig:2}, as a function of 
S\'ersic index, $n$, and intrinsic axis ratio, $\qint$.\footnote{Again, we define the projected 
effective radius $\Reff=a$ as this lies in the plane of axisymmetry 
--- assumed to be the rotation midplane --- which is the projected major axis for oblate cases, 
but the projected minor axis for prolate cases.} 
\citealt{vandeVen21} make a similar comparison for both axisymmetric and triaxial systems using an approximation 
for $\rho$, but show \rhalftD relative to the projected major axis, so for the axisymmetric, 
prolate systems our ratio differs from theirs. 
The 3D spherical half-mass radius is larger than the 2D projected effective radius 
enclosing half of the total light (and half of the total mass, assuming constant $M/L$ and an optically thin medium). 
There is a larger dependence of the ratio $\rhalftD/\Reff$ on \qint than on $n$, 
where $\rhalftD/\Reff\sim1.3-1.36$ when $\qint=1$ for all $n=0.5-8$, 
but as $\qint$ decreases the ratio decreases towards $\rhalftD/\Reff\sim1$ for all $n$.

Next, we examine how the relation between the mass and circular velocity profiles 
deviates from the relation that holds for spherical symmetry, where $\Menc(<r=R) = \vcirc^2(R) R / G$. 
In Figure~\ref{fig:3} we show computed fractional enclosed mass (top) 
and circular velocity profiles (bottom) for $n=1,4$ (top and bottom rows, respectively) and 
for $\qint=1,0.4, 0.2$ (left, center, right columns, respectively).

For the spherically symmetric ($\qint=1$) cases, the numerical evaluation of $\Menc(<r=R)$ (red dashed line; Eq.~\ref{eq8}) 
and $\vcirc(R)$ (black dotted line; Eq.~\ref{eq5}) follow the expected relation ($\Menc(<r=R) = \vcirc^2(R) R / G$), 
and the isodensity spheroids are spherical, so there is no difference between the enclosed spherical and spheroidal profiles. 
Echoing the previous figures, we also see that the enclosed 3D mass profile 
increases more slowly as a function of $R$ than the 2D projected profile (solid orange line; Eq.~\ref{eq3}).

In contrast, for flattened deprojected models with $\qint<1$,  the deviation of the $\Menc(<r=R)$ and 
$\vcirc(R)$ profiles from the spherical relation become more pronounced for lower intrinsic axis ratios. 
Also, $\sqrt{\Mencspheroid(<m=R)G/R}$ (purple dashed-dotted line; Eq.~\ref{eq6}) does not match the correct $\vcirc(R)$ curve. 
As $\qint$ decreases, $\Menc(<r=R)$ approaches the projected 2D ellipse curve, because for flatter deprojected models 
there is less additional mass outside the sphere along the remaining line-of-sight collapse.

\begin{figure*}[!ht]
\vspace{-2pt}
\centering
\hglue -5pt
\includegraphics[width=1.02\textwidth]{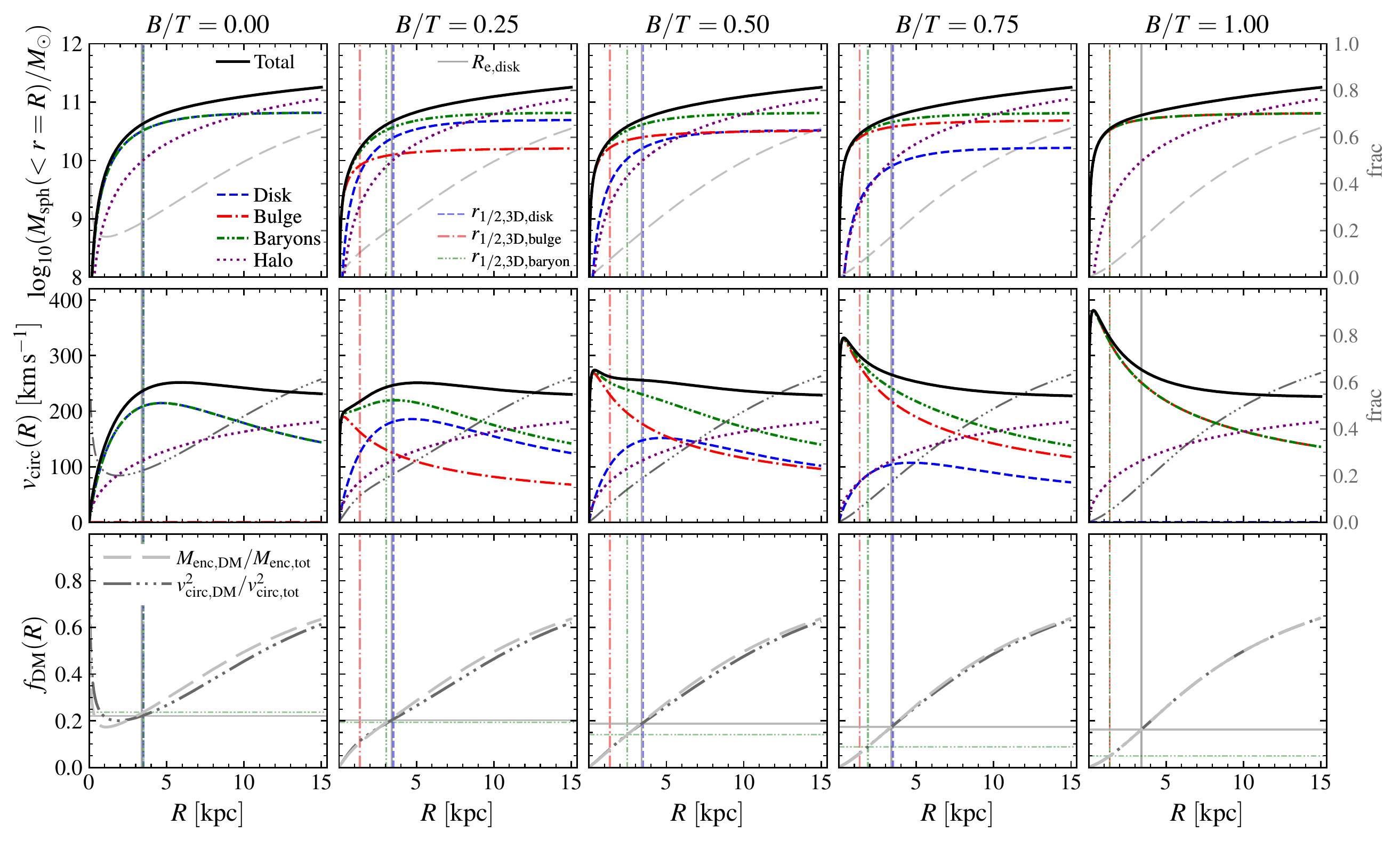}
\caption{Enclosed mass (3D spherical, \emph{top}), circular velocity (\emph{middle}), and dark matter (\emph{bottom})
profiles for different components of an example galaxy as a function of projected major axis radius, 
for bulge-to-total ratios of $\bt=0, 0.25, 0.5, 0.75, 1$ (\emph{left to right}). 
For all cases, we compute the mass components assuming values for 
a typical $z=2$ massive main-sequence galaxy with $\lMstar=10.5$:
$\Mbar=6.6\times10^{10}\Msun$, $\Redisk=3.4\,\unit{kpc}$, 
$\nSdisk=1$, $\qintdisk=0.25$, $\Rebulge=1\,\unit{kpc}$, 
$\nSbulge=4$, $\qintbulge=1$, and a NFW halo with $\Mhalo=8.9\times10^{11}\Msun$ and 
$c=4.2$. 
Shown are the $\Menc(<r=R)$ and $\vcirc(R)$ profiles for the 
disk (dashed blue), bulge (dash-dot red), total baryons (disk+bulge; dash-dot-dot green), halo (dotted purple), 
and composite total system (solid black). 
Vertical lines mark $R=\Redisk$ (solid grey) and the 3D spherical half-mass radii \rhalftD 
for the disk (dashed blue), bulge (dash-dot red), and total baryons (dash-dot-dot green). 
Two dark matter fraction definitions are shown in the bottom panels, 
$\fDMm=\MencDM/\Menctot$ and $\fDMv=\vcircDM^2/\vcirctot^2$, 
with long dashed grey and long dash-triple-dotted dark grey lines, respectively. 
(The $\fDMm$ and $\fDMv$ curves are also shown in the top and middle panels, 
respectively, with the scale at the right axis of each panel.) 
When a disk component is present, the system is no longer spherically symmetric, 
so $\MencDM/\Menctot$ and $\vcircDM^2/\vcirctot^2$ differ. 
This deviation is larger when the disk contribution is large (i.e., lower \bt), 
though even at low \bt the difference is relatively modest (see also Figure~\ref{fig:6}). 
Additionally, while the ratio $\rhalftD/\Reff$ for a single component (e.g., the disk or bulge) is generally modest 
(see Figure~\ref{fig:2}), for a composite disk+bulge system the total baryon $r_{\mathrm{1/2,3D,baryon}}$ 
becomes much smaller relative to \Redisk with increasing \bt (vertical green dash-dot-dot and solid grey lines). 
If such disparate ``half'' radii definitions are used to define \fDM apertures 
(i.e., $\fDMv(\Redisk)$ versus $\fDMm(r_{\mathrm{1/2,3D,baryon}})$, horizontal solid grey and green dash-dot-dot lines), 
this leads to increasingly large offsets between the \fDM values towards higher \bt 
(see also Figure~\ref{fig:7}).}
\label{fig:5}
\end{figure*}

\vspace{-5pt}

\section{Virial coefficients for enclosed 3D and total masses}
\label{sec:virialcoeff}

We now quantify the relationship between mass- and velocity-derived quantities for different 
S\'ersic indices and intrinsic axis ratios. 
By including a ``virial'' coefficient $\kthrd(R)$ which depends on the geometry and mass distribution (\citealt{BT08}), 
the spherical enclosed mass and circular velocity can be related by 
\vspace{-5pt}
\begin{equation}
\Menc(<r=R)=\kthrd(R)\frac{\vcirc^2(R) R}{G}.
\label{eq9}
\end{equation} 
This virial coefficient is evaluated by combining Eqs.~\ref{eq5}~and~\ref{eq8}.

For comparison with integrated galaxy quantities, 
it is also useful to define a ``total'' virial coefficient $\ktot(R)$ 
which relates the total system mass to the circular velocity at a given radius: 
\vspace{-2pt}
\begin{equation}
\Mtot=\ktot(R)\frac{\vcirc^2(R) R}{G}.
\label{eq10}
\end{equation}

Figure~\ref{fig:4} shows 
$\ktot(R=\Reff)$ and $\kthrd(R=\Reff)$ versus S\'ersic index $n$ for a range of \qint. 
For the spherical case ($\qint=1$), $\kthrd(R=\Reff)=1$, as expected by spherical symmetry. 
However, as $\Reff<\rhalftD$ for spherical deprojected S\'ersic models, $\ktot(R=\Reff,\qint=1)\neq2$, 
but instead exceeds 2 for all $n$ (i.e., 2.933 when $n=1$).  
For oblate flattened S\'ersic deprojected models (i.e., $\qint<1$), $\ktot(R=\Reff)$ is lower than the 
$\qint=1$ case for all $n$, while prolate cases ($\qint>1$) have larger $\ktot(R=\Reff)$. 
For $\kthrd(R=\Reff)$ the trends are more complex, but for $n\gtrsim2$ the oblate (prolate) 
models all have $\kthrd(R=\Reff)<1$ ($>1$).
For reference, we also present values of $\ktot(R)$ and $\kthrd(R)$ for a range of $R$, $n$, and $\qint$ in Table~\ref{tab:virialcoeff}.
These total virial coefficients in particular allow for a more precise comparison between 
the dynamical \Mtot and projection-derived quantities, such as \Mstar or \Mgas, 
particularly when full dynamical modeling is not possible (e.g., the approach used 
in \citealt{Erb06c}, \citealt{Miller11}, \citealt{Price16,Price20}, and numerous other studies).


\section{Mass distributions of multi-component galactic systems}
\label{sec:mdistgal}

\vspace{-2pt}

\subsection{Mass and velocity distributions of systems including both flattened and spherical components}
\label{sec:massfractiondistro}

While the virial coefficients derived in the previous section allow for the conversion from 
circular velocities to enclosed masses for a single non-spherical mass component, 
observed galaxies tend to have multiple mass components, of varying intrinsic shapes and profiles. 
We now explore the enclosed mass and circular velocity distributions for galaxies with multiple mass components, 
focusing on how the non-spherical components impact the  mass fraction distributions inferred from velocity profile ratios.

We calculate profiles for a ``typical'' $z=2$ main-sequence star-forming galaxy of stellar mass $\lMstar=10.5$ 
consisting of a bulge, disk, and halo, over a range of bulge-to-total ratios, \bt. 
We thus adopt a total $\Mbar=6.6\times10^{10}\Msun$ 
(using the gas fraction scaling relation of \citealt{Tacconi20}). 
We assume a thick, flattened disk modeled as a deprojected S\'ersic distribution 
with $\qintdisk=0.25$, and adopt $\nSdisk=1$, 
$\Redisk=3.4\unit{kpc}$ 
(from observed trends and scaling relations; \citealt{Wuyts11b}, \citealt{vanderWel14a}). 
The bulge is modeled as a deprojected S\'ersic component with $\nSbulge=4$, $\Rebulge=1\unit{kpc}$, and $\qintbulge=1$. 
We also include a NFW halo without adiabatic contraction, assuming 
$\mathrm{conc}_{\mathrm{halo}}=4.2$  and $\Mhalo=8.9\times10^{11}\Msun$ 
(following observed halo concentration and stellar mass to halo mass relations, 
e.g. \citealt{Dutton14}, \citealt{Moster18}).\footnote{However, 
many massive SFGs at these redshifts exhibit lower \fDM that suggest 
more cored halo profiles; see e.g. \citealt{Genzel20}.}
In Figure~\ref{fig:5}, for each of the \bt ratios (left to right), we show the enclosed mass (top row) 
and circular velocity profiles (middle row) as a function of radius. 
The impact of shifting the baryonic mass from entirely in the disk (the only oblate, non-spherical mass component; $\bt=0$) 
to entirely in the bulge (only spherical mass components; $\bt=1$) can be seen in both the \Menc and \vcirc profiles. 
The lower S\'ersic index and larger \Reff of the disk (blue dashed line) relative to the bulge (red dash-dot)
result in more slowly rising mass and \vcirc curves for the baryonic component (green dash-dot-dot) at low \bt, 
with the curves rising more quickly as the bulge contribution increases. 
The total galaxy mass and \vcirc curves (solid black) are dominated by the baryonic components 
in the inner regions, but at large radii ($R\gtrsim10\unit{kpc}$) 
where the halo begins dominating the mass and \vcirc profiles, the curves are similar for all \bt.

We also show the radial variation of the  3D enclosed halo to total mass ratio $\MencDM/\Menctot$ (long light grey dashed line)
and the squared halo to total circular velocity ratio $\vcircDM^2/\vcirctot^2$ (dark grey dash-triple dot line) 
in the bottom row (and the \Menc and \vcirc panels, respectively). 
As expected when $\bt<1$, these two ratios are not equivalent, though they become closer as $\bt\to1$ and 
more of the total galaxy mass is found in spherical components. 
For $\bt=1$, the galaxy is spherically symmetric, so the two ratios are equal.

\begin{figure}
\vspace{-6pt}
\centering
\includegraphics[width=0.49\textwidth]{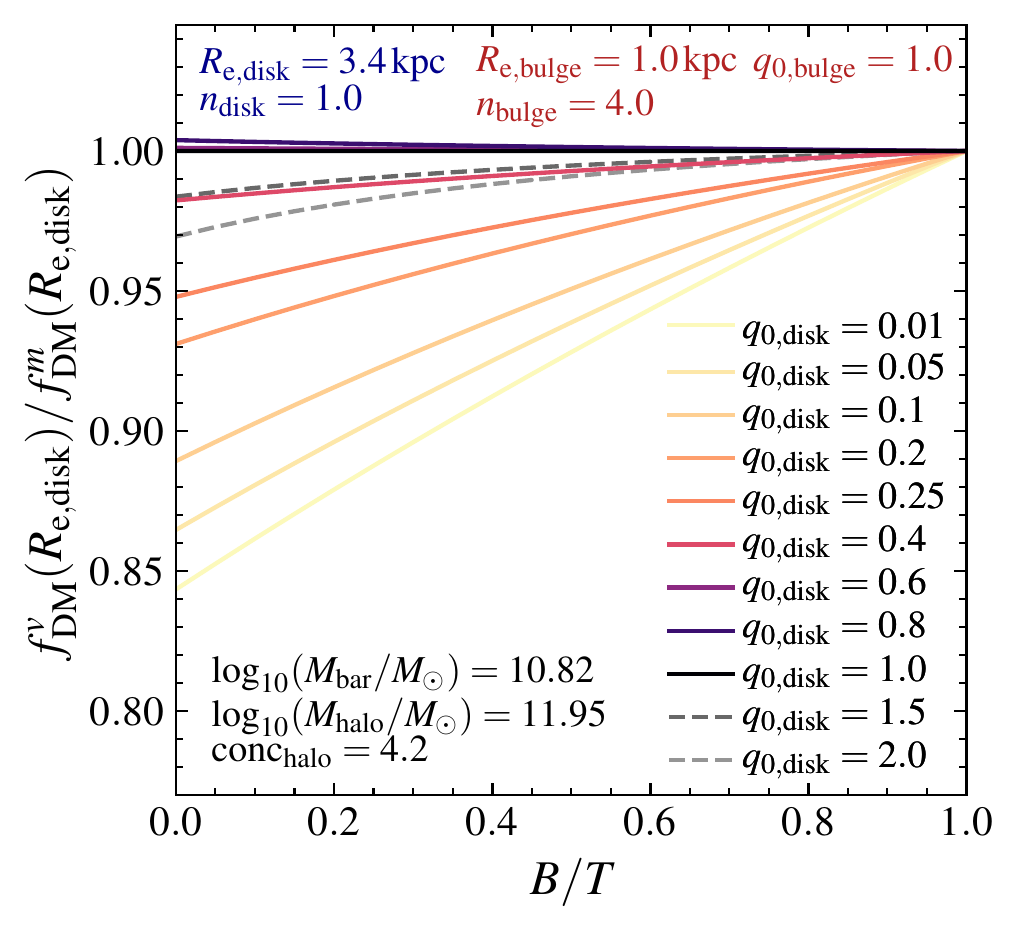}
\vspace{-20pt}\caption{Ratio between the dark matter fraction at \Redisk calculated from the 
circular velocity and from the 3D spherical enclosed mass, $\fDMv(\Redisk)/\fDMm(\Redisk)$, 
versus bulge-to-disk ratio \bt, for a range of different disk intrinsic axis ratios (colored lines, from $\qintdisk=0.01$ to 1) 
for an example massive galaxy at $z=2$. 
The ratio between the two dark matter fraction measurements is lower for lower \bt (i.e., higher disk contributions) 
and lower \qintdisk (i.e., more flattened disks), with $\fDMv(\Redisk)/\fDMm(\Redisk)\lesssim 0.9$ 
for low values of both \qintdisk and \bt.
The limiting case of a Freeman (infinitely thin) exponential disk has $\fDMv(\Redisk)/\fDMm(\Redisk)\approx 0.84$. 
As \bt increases for fixed \qintdisk, and likewise for increasing \qintdisk at fixed \bt, 
the ratio of the two fraction measurements approaches 1 because the composite system becomes more spherical. 
Overall, the discrepancy between the \fDM estimators measured at the same radius is relatively minor.}
\label{fig:6}
\vspace{-6pt}
\end{figure}

\vspace{-5pt}

\subsection{Defining dark matter fractions}
\label{sec:deffDM}

As illustrated in Figure~\ref{fig:5}, the approximation $v^2_{\mathrm{circ,DM}}(R)/v^2_{\mathrm{circ,tot}}(R)$ 
deviates from the enclosed spherical mass fraction $\MencDM(<r=R)/\Menctot(<r=R)$ 
for galaxies with a non-spherical disk component, particularly when the \bt ratio is $\lesssim0.5$. 
This deviation thus leads to differences in inferred dark matter fractions, 
depending on how the fraction is defined.

If the dark matter fraction is defined as the ratio of the dark matter to total mass enclosed within a sphere of a given radius, 
we have $\fDMm(R)=\MencDM(<r=R)/\Menctot(<r=R)$. 
This approach is often adopted for simulations, where it is easy to determine mass within a given radius. 
However, observations cannot directly probe the mass distributions, so generally the fraction is defined 
based on the circular velocity ratio, $\fDMv(R)=v_{\mathrm{circ,DM}}^2(R)/\vcirctot^2(R)$. 
If a galaxy has only spherically-symmetric components, these two definitions are equivalent 
(as seen in the right column of Figure~\ref{fig:5}), but as noted in Figure~\ref{fig:5}, 
the two definitions are no longer equivalent with non-spherical components, where \fDMm is generally larger than \fDMv.

To further quantify how much these definitions can vary, we compare the value of the ratio between \fDMv and \fDMm 
at \Redisk over a range of \bt ratios and intrinsic disk thicknesses \qintdisk in Figure~\ref{fig:6}. 
For this example case (using a massive galaxy with $\Mbar=6.6\times10^{10}\Msun$ at $z=2$, 
as in Figure~\ref{fig:5}), 
we see that $\fDMv(\Redisk)$ can be as low as $\sim85\%$ of $\fDMm(\Redisk)$ (in the extreme case 
with $\qintdisk=0.01$). For more typical expected disk thicknesses for 
galaxies at $z\sim1-3$, $\qintdisk\sim0.2-0.25$, 
we find a minimum of $\fDMv(\Redisk)/\fDMm(\Redisk) \sim 0.93-0.95$ (for $\bt=0$). 
While this deviation is fairly small in this example, using consistent definitions of \fDM when comparing 
simulations and observations would avoid introducing systematic shifts between the values.

\begin{figure*}[ht!]
\vspace{-6pt}
\centering
\hglue -5pt
\includegraphics[width=1.02\textwidth]{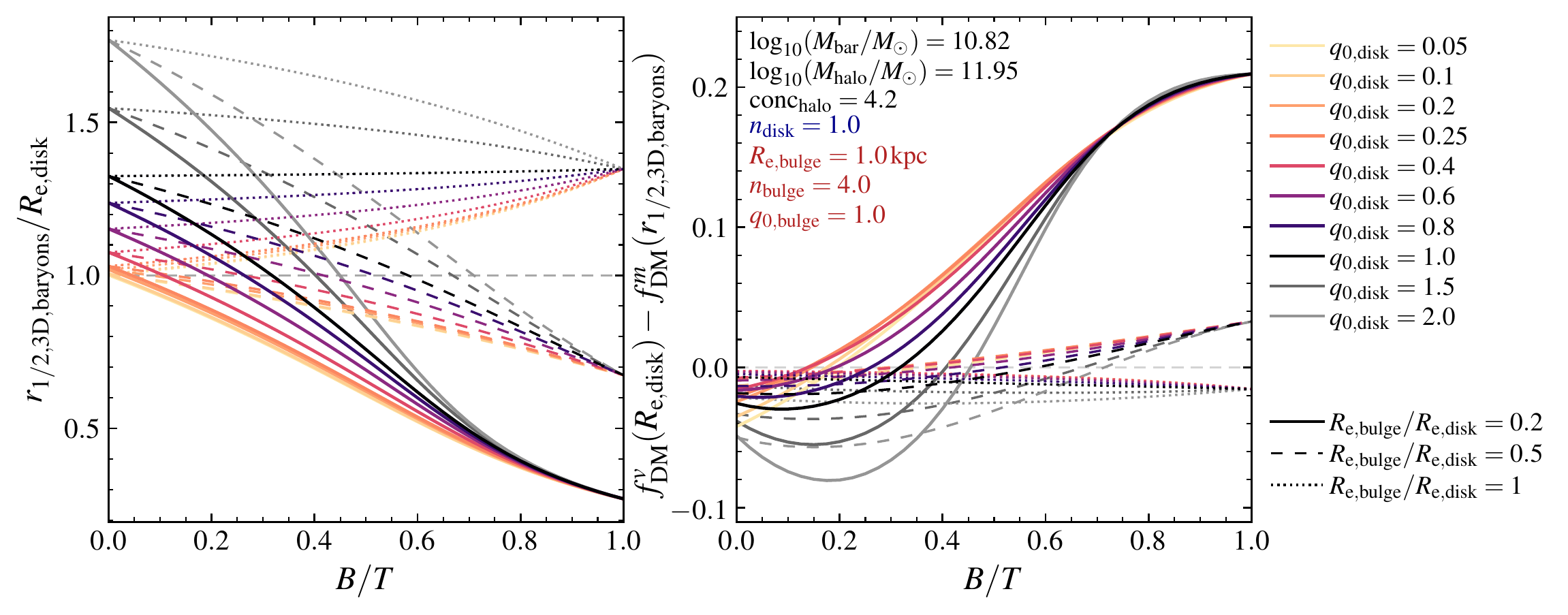}
\vspace{-20pt}
\caption{Ratio between the composite disk+bulge 3D half-mass radius 
and the 2D projected disk effective radius ($\rhalftDbar/\Redisk$; \emph{left}) 
and the difference between the dark matter fraction estimators at these radii 
($\fDMv(\Redisk)-\fDMm(\rhalftDbar)$; \emph{right}), as a function of \bt, 
for a range of disk intrinsic axis ratios (colored lines, from $\qintdisk=0.05$ to 2) 
and ratio between the bulge and disk \Reff (solid, dashed, and dotted lines, for 
$\Rebulge/\Redisk=0.2,0.5,1$, respectively). 
The adopted galaxy values are the same as in Figure~\ref{fig:6}, 
except \Redisk is now determined by $\Rebulge/\Redisk$. 
With a non-zero bulge contribution, $\rhalftDbar/\Redisk$ deviates 
from the single-component ratio (Figure~\ref{fig:2}), 
decreasing with increasing \bt for $\Redisk=2,5\unit{kpc}$ 
for all \qintdisk (though increasing with \bt when $\qint<1, \Redisk=\Rebulge=1\unit{kpc}$). 
For large \bt and $\Redisk=5\unit{kpc}$, the composite \rhalftDbar is less than 50\% of \Redisk. 
If the dark matter fractions are measured at different radii, 
the mismatch of the aperture sizes will lead to much larger 
\fDM differences than those found for the simple estimator mismatch 
(\fDMm vs \fDMv at the same radius; Figure~\ref{fig:6}). 
Here, we have shown $\fDMv(\Redisk)$, as might be adopted for modeling of observations, 
and $\fDMm(\rhalftDbar)$, representing a simple option for simulations 
(where spherical curves of growth separating gas, star, and DM particles could be used 
to find both the composite baryon \rhalftDbar within, e.g., $R_{\mathrm{vir}}$ and then \fDMm). 
For small \bt, $\fDMm(\rhalftDbar)$ is larger than $\fDMv(\Redisk)$, 
but for large \bt, the trend reverses (excepting the $\Redisk=\Rebulge=1\unit{kpc}$ case), 
and $\fDMv(\Redisk)$ can be up to 50\%-400\% larger than $\fDMm(\rhalftDbar)$ 
as $\bt\to1$ (for $\Redisk=2,5\unit{kpc}$, respectively). 
This example illustrates how, depending on galaxy structures, 
quoted ``half-mass'' \fDM values can be very different --- 
but that this is primarily driven by the aperture radii definitions and not by estimator mismatches.}
\label{fig:7}
\end{figure*}

\begin{figure*}[th!]
\centering
\includegraphics[width=0.91\textwidth]{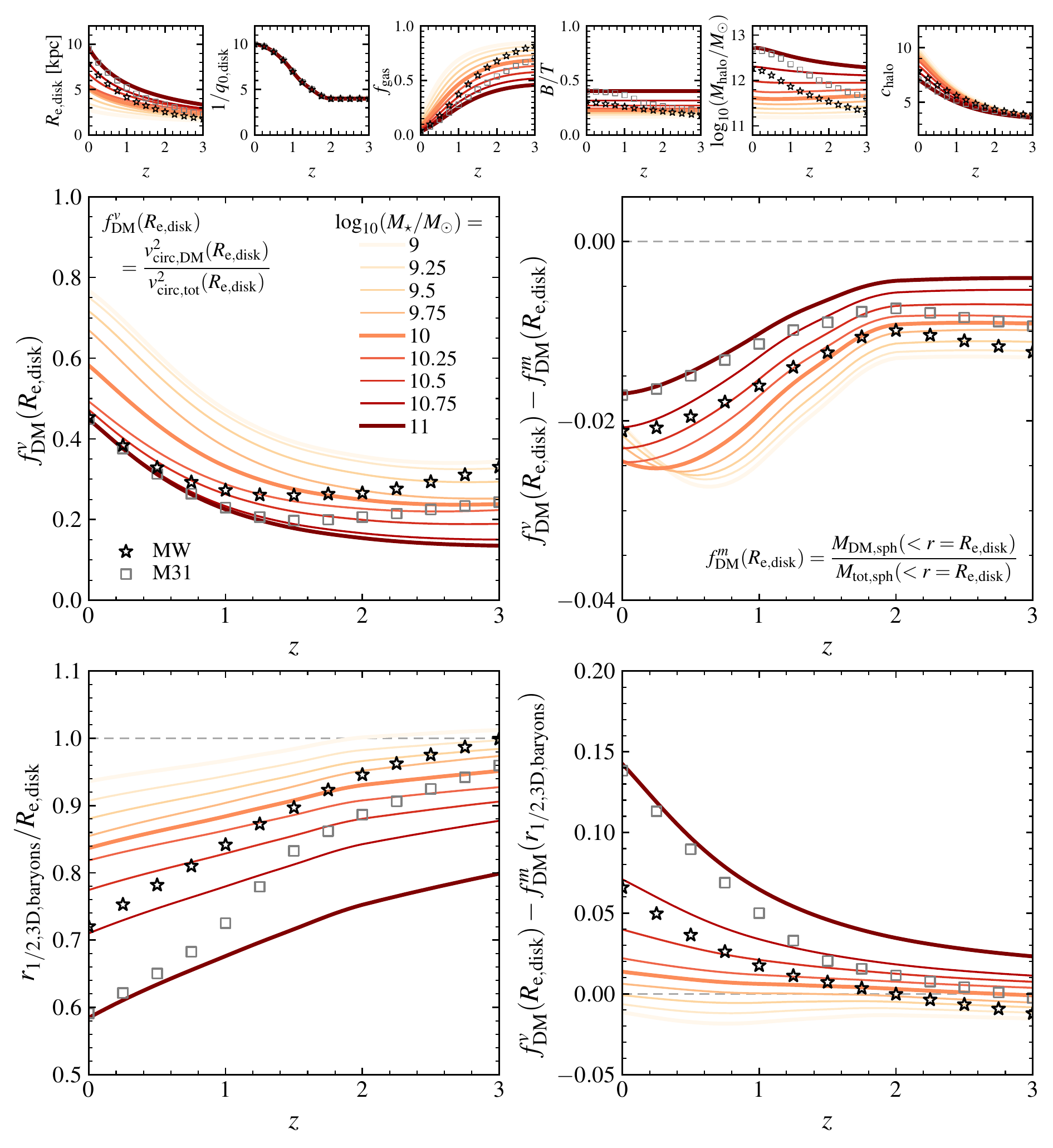}
\vglue -2pt
\caption{Toy model of how $\fDMv(\Redisk)$ (\emph{upper left}), $\fDMv(\Redisk)-\fDMm(\Redisk)$ (\emph{upper right}), 
 $\rhalftDbar/\Redisk$ (\emph{lower left}), and 
 $\fDMv(\Redisk)-\fDMm(\rhalftDbar)$ (\emph{lower right}) 
vary with redshift for a range of fixed \lMstar, 
using ``typical'' galaxy sizes, intrinsic axis ratios, gas fractions, \bt ratios, halo masses, and halo concentrations 
(from empirical scaling relations or other estimates; \citealt{Dutton14}, \citealt{Lang14}, 
\citealt{vanderWel14a}, \citealt{Moster18}, \citealt{Ubler19}, \citealt{Tacconi20}, \citealt{Genzel20}). 
The assumed (interpolated, extrapolated) property profiles as a function of redshift for each of the fixed 
\lMstar are shown in the top panels. Using abundance-matching models (inferred from Fig.~4 of \citealt{Papovich15}, 
based on the models of \citealt{Moster13}), we show the path of a Milky Way (MW, $\Mstar=5\times10^{10}\Msun$ at $z=0$;  black stars) 
and M31 progenitor ($\Mstar=10^{11}\Msun$ at $z=0$; grey squares) 
over time in each of the panels, assuming the progenitors are ``typical'' at all times. 
This inferred ``typical'' evolution would predict an increase in $\fDM(\Redisk)$ with time at fixed \Mstar, 
with lower masses having higher \fDM at all $z$. 
The evolution of the structure and relative masses of the disk, bulge, and halo 
predict an increase ($\Mstar\gtrsim10^{10.25}\Msun$) or ``dip'' ($\Mstar\lesssim^{10.25}\Msun$) 
in the difference $\fDMv(\Redisk)-\fDMm(\Redisk)$ 
between $z\sim0$ and $z\sim0.75$, and then an increase until $z\sim2$ when the 
difference flattens (largely reflecting the flat \qintdisk estimate for $z\gtrsim2$). 
The difference is minor, between $\sim-0.025$ and $-0.005$ for the stellar masses shown. 
The ratio of the composite $\rhalftDbar/\Redisk$ increases with redshift for all masses, 
with more massive models predicting smaller ratios at each $z$. 
The MW and M31 progenitors have $\fDM(\Redisk)$ evolving from $\sim0.33$ and $\sim0.25$ (respectively) at $z=3$, 
decreasing to $\sim0.25$ and $\sim0.2$ at $z\sim1.5$, and then increasing to roughly same value $\sim0.45$ at $z=0$. 
The $\fDMv(\Redisk)$ and $\fDMm(\rhalftDbar)$  values are relatively similar down to $z\sim1.5$, 
but at lower redshifts (where $\rhalftDbar/\Redisk\lesssim0.9$) 
the difference increases up to $\sim0.065$ (MW) and $\sim0.14$ (M31) at $z\sim0$. 
While this ``typical'' case predicts \fDM offsets of only 0.035 at $z=2$ and increasing to 0.14 at $z\sim0$ for the most massive case, 
objects with even larger bulges ($\bt>0.4$) or radii above the mass-size relation will have 
even more discrepant \fDM values when adopting these radii definitions (see Figure~\ref{fig:7}).}
\label{fig:8}
\vglue -30pt
\end{figure*}

\vspace{-5pt}

\subsection{Impact of aperture effects on dark matter fractions}
\label{sec:fDMaper}

While the fractional differences between the 3D half mass radii \rhalftD and the projected 2D effective radii \Reff 
for a single component, and between the \fDMv and \fDMm definitions are generally small for 
expected galaxy thicknesses, measuring \fDM (of either indicator) at \emph{different} radii --- 
such as the easily measurable \rhalftD for simulations versus \Reff for observations ---
can lead to very large discrepancies in the \fDM values. 
We demonstrate this issue in Figure~\ref{fig:7}.

First, while we show the ratio of \rhalftD/\Reff for a single component in Figure~\ref{fig:2}, 
the ratio for a multi-component system is not self-similar, but depends also on the ratio of effective radii for the components. 
For a disk + bulge system, a number of observational studies use the disk effective radius as the 
dark matter fraction aperture. We thus determine the 3D half-mass radius for the composite disk+bulge system, 
and plot the ratio \rhalftDbar/\Redisk as a function of \bt in the left panel of Figure~\ref{fig:7}, 
for a range of ratios \Rebulge/\Redisk (line style) and disk intrinsic thicknesses (\qintdisk, assuming a spherical bulge). 
Depending on the \Rebulge/\Redisk and \qintdisk values, this ratio can range from $\sim0.3-1.3$ 
(for oblate or spherical disk geometries, or up to $\sim1.7$ for prolate disks), 
with the lowest values arising from the combination of a low \Rebulge/\Redisk and a high \bt.

We then demonstrate the effects of measuring \fDM at these different aperture radii in the right panel of 
Figure~\ref{fig:7}. 
Here we plot the absolute difference $\fDMv(\Redisk)-\fDMm(\rhalftDbar)$ as a function of \bt, calculated 
for the same \Rebulge/\Redisk and \qintdisk values.  
For consistency, the \fDM estimators for each are chosen to reflect the typical definitions from observations and simulations, 
respectively, in line with the ``half-mass'' radii choices (though, as seen in Figure~\ref{fig:6}, using 
\fDMv versus \fDMm contributes very little to the differences seen in this figure). 
For very low \bt, most cases produce $\fDMv(\Redisk)-\fDMm(\rhalftDbar) \sim-0.025$ 
(e.g., larger $\fDMm(\rhalftDbar)$). 
For most practical cases with a larger disk than bulge ($\Rebulge/\Redisk<1$), 
the difference increases towards larger \bt, with $\fDMv(\Redisk)>\fDMm(\rhalftDbar)$ 
by $\bt\sim0.2-0.5$ (for $\Rebulge/\Redisk=0.2,0.5$, respectively). 
This difference can be very large, up to $\sim0.2$ for large \bt and low $\Rebulge/\Redisk$, 
as might be expected for massive galaxies that simultaneously have 
massive bulges (i.e., high \bt) but also large disk effective radii (i.e., low \Rebulge/\Redisk).

We extend these test cases to consider how \fDM for the different definitions and apertures 
change with redshift and stellar mass for a ``typical'' star-forming galaxy, as shown in Figure~\ref{fig:8}. 
We use empirical scaling relations or other estimates to determine \Redisk, \qintdisk, \fgas, \bt, \lMhalo, and \chalo 
for a range of $z$ and \lMstar (assuming the disk and bulge follow deprojected S\'ersic models, 
with fixed $\nSdisk=1$, $\nSbulge=4$, and $\Rebulge=1\unit{kpc}$; top panels). 
These toy models predict $\fDMv(\Reff)$ (Figure~\ref{fig:8}, center left) 
to increase over time at fixed stellar mass (in part because of the increasing \chalo and \Redisk over time), 
and that lower \Mstar galaxies have higher $\fDMv(\Reff)$ at fixed redshift, with 
relatively low $\fDMv(\Reff)$ for the most massive galaxies ($\sim20\%$) at $z\sim1-3$. 
This is qualitatively in agreement with recent observations (e.g., \citealt{Genzel20}, \citealt{Price21}, \citealt{Bouche22}), 
though these recent studies also provide evidence for non-NFW halo profiles (in particular, cored profiles), 
which would produce lower $\fDMv(\Reff)$ for the same \Mhalo than our toy models. 
As a further example, we consider how the predictions would change over time for a Milky Way and M31-mass progenitor. 
In both cases, the predicted $\fDMv(\Reff)$ decrease from $z=3$ to a minimum at $z\sim1.5$, 
and then increase until the present day. 
For these toy values, the difference in dark matter fraction definitions measured \emph{at the same radius} 
(\Redisk; Figure~\ref{fig:8}, center right)  
typically differ by only $\sim-0.005$ to $\sim-0.025$ (typically $\sim4-6\%$ of the measured values), 
with a larger typical offset at lower redshifts and for lower masses.

We find the ratio of the 3D baryonic half mass radius to the disk effective radius 
($\rhalftDbar/\Redisk$; Figure~\ref{fig:8}, bottom left) 
for these models decreases towards lower redshifts and with increasing stellar mass, 
from \hbox{$\sim1$} at \hbox{$z\sim2-3$} to \hbox{$\sim0.94$} at \hbox{$z=0$} for the lowest \Mstar, 
and from \hbox{$\sim0.8$} at \hbox{$z=3$} down to \hbox{$\sim0.6$} at \hbox{$z=0$} for the highest \Mstar. 
For the MW and M31 progenitors, this ratio decreases from \hbox{$\sim1,0.96$} at \hbox{$z=3$} 
down to \hbox{$\sim0.72,0.59$} at \hbox{$z=0$}, respectively. 
The difference between the two dark matter fraction definitions measured at these different radii apertures 
($\fDMv(\Redisk)-\fDMm(\rhalftDbar)$; Figure~\ref{fig:8}, bottom right) 
for these toy models is typically much larger than the difference for the definitions alone, 
and tends to increase towards lower redshifts and with stellar mass. 
The difference changes from \hbox{$\sim-0.015, 0., 0.025$} at \hbox{$z=3$} 
to \hbox{$\sim -0.01,0.015,0.14$} at \hbox{$z=0$} for \hbox{$\lMstar=9, 10, 11$}, respectively. 
The MW and M31 progenitors have offsets increasing from \hbox{$\sim-0.013, -0.003$} 
at \hbox{$z=3$} to \hbox{$\sim0.065,0.14$} at \hbox{$z=0$}, respectively.

\begin{figure*}[t!]
\vspace{-5pt}
\centering
\includegraphics[width=0.85\textwidth]{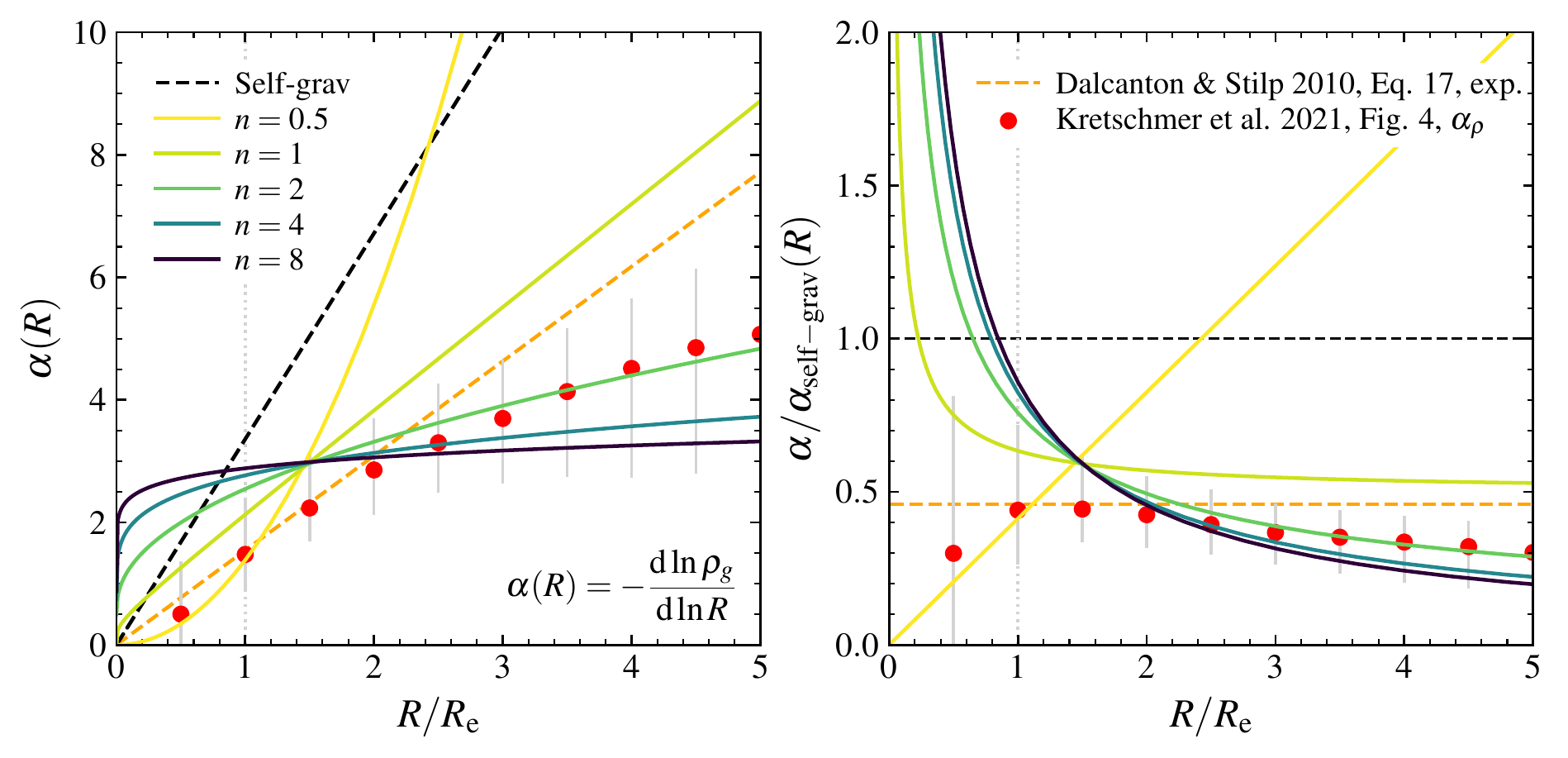}
\vspace{-6pt}
\caption{The pressure support correction, $\alpha(R)$, versus $R/\Reff$ for 
a self-gravitating exponential disk and deprojected S\'ersic models. 
The left panel directly compares $\alphaSG(r)=3.36(R/\Reff)$ for the self-gravitating disk 
(as in \citealt{Burkert10}; black dashed line) to \hbox{$\alpha(R,n)=-\D\ln\rho(R,n)/\D\ln R$}  
determined for a range of S\'ersic indices $n$ (colored lines). 
The ratio $\alpha/\alphaSG(R)$ is shown in the right panel. 
For $n\geq1$, \alphan is smaller than \alphaSG when $R\gtrsim0.2-0.8\Reff$, 
though $\alpha(n\geq1)$ does exceed \alphaSG at the smallest radii. 
This implies that for most radii, there is less asymmetric drift correction (and thus higher $v_{\mathrm{rot}}$) 
for the deprojected S\'ersic models (e.g., $n=1$) than for the self-gravitating disk. 
However, for $n=0.5$, \alphan is greater than \alphaSG at  $R\gtrsim2.4\Reff$, 
so at large radii the $n=0.5$ deprojected S\'ersic model
predicts a larger pressure support correction than for the self-gravitating disk case.  
The lower pressure support predicted for $\alpha(n\gtrsim1)$ than for \alphaSG 
is in agreement with recent predictions from simulations by \citet{Kretschmer21} (red circles; 
with the vertical grey bars denoting the $1\sigma$ distribution),
as well the relation by \citet{Dalcanton10} for a power law relationship between the 
gas surface density and the turbulent pressure (orange dashed line).}
\label{fig:9}
\end{figure*}

$\qquad$ 

$\qquad$

$\qquad$

$\qquad$

Given the very modest offsets for the \fDM definition differences alone, 
these offsets are nearly entirely driven by the differences between the aperture radii. 
Indeed, though the differences for these toy calculations --- driven almost entirely by the aperture mismatches ---
do not reach the extreme differences of 
$\fDMv(\Redisk)-\fDMm(\rhalftDbar)$ seen in Figure~\ref{fig:7} 
for parts of the parameter space (in part because the maximum toy model \bt is \hbox{$\sim0.4$}), 
we still predict absolute differences up to almost \hbox{$\sim0.15$} at \hbox{$z=0$}, 
and \hbox{$\sim0.03-0.07$} at \hbox{$z\sim1-2$}. 
This offset is on par with the current observational uncertainties at $z\gtrsim1$ (\hbox{$\sim0.1-0.2$}; e.g., \citealt{Genzel20}). 
To ensure the most direct comparison between observations and simulations --- 
particularly as observational constraints on \fDM at higher redshifts continue to improve --- 
it will be important to account for such aperture differences (either by measuring in equivalent apertures, or by applying 
an appropriate correction factor) in order to better
determine if, and how, observation and simulation predictions differ.


\section{Turbulent pressure support effects on rotation curves}
\label{sec:asymmdrift}

\subsection{Derivation of pressure support for a single component}
\label{sec:asymmdrift_single}

As many dynamical studies of high-redshift, turbulent disk galaxies use gas motions as the dynamical tracer, 
we now consider how turbulent pressure support will modify the rotation curves 
if the gas is described by a deprojected S\'ersic model. 
We follow the derivation of \citet{Burkert10}, and also assume the pressure support is due only 
to the turbulent gas motions (i.e., the thermal contribution is negligible). 
We thus begin from Eq.~2 of \citet{Burkert10}, where the pressure-corrected gas rotation velocity is 
\begin{equation}
\vrot^2(R) = \vcirc^2(R) + \frac{1}{\rho_g} \frac{\D}{\D\ln{}R} \left(\rho_g\sigma^2\right), 
\label{eq:B10-2}
\end{equation}
where $\vcirc$ is the circular velocity in the midplane of the galaxy determined from the total system potential 
(including all mass components: stars, gas, and halo; i.e., the rotational velocity if there is no pressure support), 
$\rho_g$ is the gas density, and $\sigma$ is the (one-dimensional) gas velocity dispersion. 
While the gas has the same circular velocity as the total system, 
the pressure support correction term from the turbulent gas motions only applies to the gas rotation 
and only depends on the gas density distribution and the gas velocity dispersion.

This relation can be generally rewritten as 
\begin{equation}
\vrot^2(R) = \vcirc^2(R) - \sigma^2 \alpha(R),
\label{eq:gen_AD}
\end{equation}
where $\alpha(R) = -\Big(\frac{\D\ln\rho_g}{\D\ln{}R} + \frac{\D\ln\sigma^2}{\D\ln{}R}\Big)$.
If we assume the velocity dispersion $\sigma=\sigmaint$ is constant, 
then this simplifies to
 \begin{equation}
\alpha(R) = - \frac{\D\ln\rho_g}{\D\ln{}R}.
\label{eq:AD_alpha_gen}
\end{equation} 
For a self-gravitating exponential disk, as assumed in \citet{Burkert10}, 
$\frac{\D\ln\rho_g}{\D\ln{}R} = 2\left(\!\frac{\D\ln\Sigma(R)}{\D\ln{}R}\!\right)$, 
which yields 
\begin{equation}
\alphaSG(R) = 2(R/R_d)=3.36(R/\Reff).
\label{eq:AD_SG}
\end{equation} 
\citet{Burkert16} generalized this result to a self-gravitating disk with an arbitrary S\'ersic index, 
where $\alpha_{\mathrm{self-grav},n}(R) = 2 \bn (R/\Reff)^{1/n}$.

\begin{figure*}
\centering
\hglue -4pt
\includegraphics[width=1.015\textwidth]{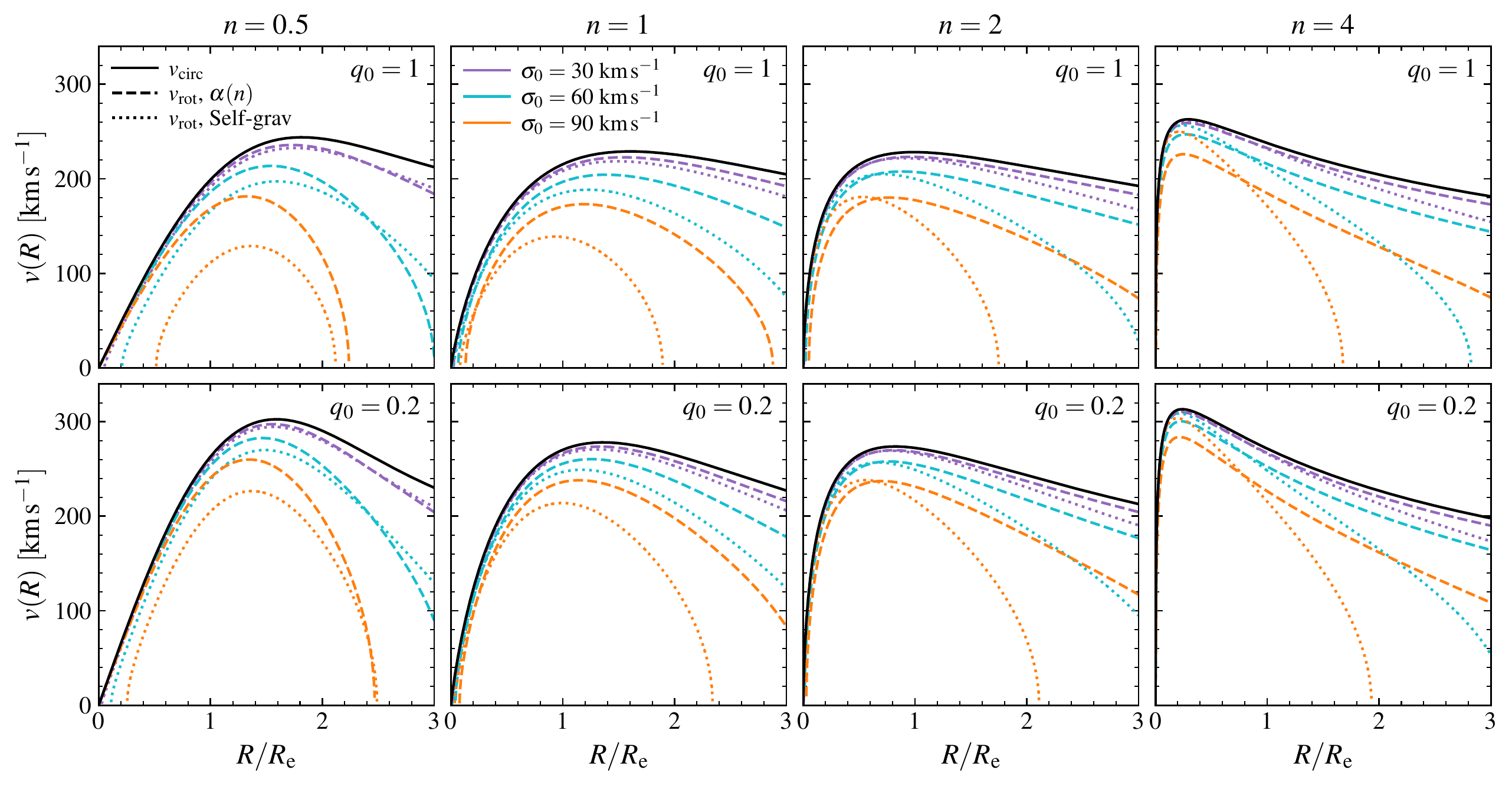}
\caption{Comparison between $v_{\mathrm{rot}}^2 = \vcirc^2-\sigmaint^2\alpha(R)$ 
determined using the deprojected S\'ersic model \alphan and the 
self-gravitating exponential disk \alphaSG (as shown in Figure~\ref{fig:9}), 
for a range of S\'ersic indices $n$, intrinsic axis ratios \qint, and velocity dispersions \sigmaint. 
For all cases, we consider a single deprojected S\'ersic mass distribution with $\Mtot=10^{10.5}\,M_{\odot}$. 
The columns show curves for $n=0.5,1,2,4$ (\textit{left to right}, respectively), while the rows show the case of 
spherical ($\qint=1$; \textit{top}) and flattened ($\qint=0.2$; \textit{bottom}) S\'ersic distributions. 
For each panel, the solid black line shows the circular velocity $\vcirc$ (determined following Eq.~\ref{eq5}). 
The colored lines show $v_{\mathrm{rot}}$ determined using \alphan (dashed) and \alphaSG (dotted), 
with the colors denoting $\sigmaint=[30,60,90]\unit{km\,s^{-1}}$ (purple, turquoise, orange, respectively). 
As expected by the $\alpha(R)$ trends shown in Figure~\ref{fig:9}, 
for $n\geq1$ we see that for most radii, the pressure support implied by \alphaSG 
results in lower $v_{\mathrm{rot}}$ than for \alphan (though at the smallest radii the inverse holds). 
In some cases, the magnitude of \sigmaint combined with the form of $\alpha(R)$ 
additionally predict disk truncation within the range shown, 
though truncation generally occurs at smaller radii for \alphaSG than for \alphan.}
\label{fig:10}
\end{figure*}

Alternatively, as derived by \citet{Dalcanton10} (their Eqs.~16 \& 17), 
for a disk with turbulent pressure $P_{\mathrm{turb}}\propto\Sigma^{0.92}$ 
(where the authors infer the exponent using results from hydrodynamical simulations of turbulence 
in stratified gas by \citealt{Joung09} combined with a Schmidt law of slope $N=1.4$; \citealt{Kennicutt98}), 
the pressure support is described by 
\begin{align}
\alpha_{\mathrm{DS10}}(R) &= -0.92\frac{\D\ln\Sigma(R)}{\D\ln{}R}=0.92 \left(\frac{\bn}{n}\right) \left(\frac{R}{\Reff}\right)^{1/n},
\label{eq:AD_DS10}\\
&= 1.5456(R/\Reff) \qquad \mathrm{for}\ n=1, 
\label{eq:AD_DS10_exp}
\end{align} 
for arbitrary $\sigma_{R}(R)$ (not only constant \sigmaint as considered here). 
Further forms of the pressure support have also been explored, as compared and discussed by \citet{Bouche22}, 
including the case for constant disk thickness (\citealt{Meurer96}, \citealt{Bouche22}), 
or when accounting for the full Jeans equation (\citealt{Weijmans08}).

For gas following a deprojected S\'ersic model, we find $\alpha(R)=\alpha(R,n)$ 
by differentiating $\rho_g=\rho(m=R,n)$.\footnote{As we are considering 
only the midplane derivative with $z=0$, $\alpha(R,n)$ is the same regardless of $\qint$.}
After combining Equations~\ref{eq2} \& \ref{eq4}, performing a change of variable, and applying the Leibniz rule, we can write 
\begin{align}
\frac{\D \rho(m)}{\D m} &=  \frac{\Upsilon}{\pi}  \frac{\qobs}{\qint} \frac{I_e \bn}{n\Reff} \frac{m}{\Reff^2} \int_0^{\infty} f(m,x) \: \D x , \label{eq17} \\ 
f(m,x) &= \exp\left\{ -\bn \left[ v^{1/n}-1\right]\right\} v^{\left(1/n -4\right)} 
\left[\frac{1}{n}- 2 - \frac{\bn}{n} v^{1/n}\right],  \nonumber \\
&\mathrm{for\ } v = \frac{1}{\Reff}\sqrt{x^2+m^2}. \nonumber
\end{align}

This expression for $\D\rho(m)/\D{}m$ can be evaluated numerically, and 
together with the numerical evaluation of $\rho(m)$, 
we have 
$\D\ln\rho/\D\ln m = (m/\rho)(\D \rho/\D m)$.\footnote{We note that 
in the limit $m\to0$, $\frac{\D\ln\rho}{\D\ln m} \to \frac{1}{n} - 1$ $\ $for $n\geq1$, 
which is helpful as numerical evaluations can be problematic at very small radii, 
particularly as the density profiles diverge at small radii when $n\geq1$.} 
Alternatively, the log density can be differentiated numerically. 
(A similar derivation of the pressure support for spherical deprojected S\'ersic profiles is 
presented in Sec.~2.2.3 of \citealt{Kretschmer21}, 
who also showed \alphan versus $n$ at select radii in their Fig.~6, 
and gave an approximate equation for \alphan at select radii in Sec.~3.5).\footnote{Note that the 
pressure correction term \alphan discussed here is the same as $\alpha_{\rho}$ as defined in \citet{Kretschmer21}. 
However, we emphasize that it is not directly comparable to the $\alpha_v$ derived by 
\citeauthor{Kretschmer21} for their simulations.  
\citeauthor{Kretschmer21} determine circular velocities from mass enclosed within a sphere, $V_c(r)=\sqrt{GM(<r)/r}$, 
and instead fold the effects of non-spherical potentials into the correction term $\Delta_Q$. 
Here we explicitly consider $\vcirc$ determined for non-spherical deprojected S\'ersic profiles, 
so \alphan does not need such a correction. 
Of course, the total $\alpha$ considered here would be modified by terms 
incorporating variable $\sigma(R)$ or anisotropic velocity dispersion, 
but these terms vanish as we assume a constant \sigmaint.}

Figure~\ref{fig:9} (left panel) shows the $\alpha(R,n)$ derived for the deprojected S\'ersic models 
as a function of radius for a range of S\'ersic index $n$ (colored lines). 
For comparison, we also show the self-gravitating disk case $\alphaSG(R)$ as presented in \citet{Burkert10} (black dashed line), 
as well as $\alpha$ determined following \citet{Dalcanton10}, 
and as measured from simulations in \citet{Kretschmer21} 
(where the density is determined from the smoothed cumulative mass profile of the cold gas, and 
\Reff of the cold gas is the half-mass radius measured within $0.1R_{\mathrm{vir}}$; c.f. Sec~2.3 \& 3.2 of \citealt{Kretschmer21}). 
The right panel additionally shows the ratio $\alpha/\alphaSG$. 
We find that \alphan is lower than \alphaSG at $R\gtrsim0.2-0.8\Reff$ for $n\gtrsim1$. 
However, at small radii ($R\lesssim\Reff$) we find $\alphan>\alphaSG$ for $n\gtrsim1$ 
(with the cross-over radius varying with $n$). 
In contrast, we find the inverse for $n=0.5$: 
$\alpha(n=0.5)$ is lower than \alphaSG up to $R\sim2.4\Reff$, but then \alphan exceeds \alphaSG at larger radii. 
In comparison to the self-gravitating disk case, we find the deprojected S\'ersic \alphan are in 
better agreement with the pressure support for an exponential distribution from \citet{Dalcanton10} 
(roughly half as much pressure support as the self-gravitating case), 
as well as with the simulation-derived pressure support by \citet{Kretschmer21} (and similar findings by \citealt{Wellons20}).
Furthermore, as demonstrated by \citet{Bouche22} (using an example \vcirc with $n=1.5$ and an NFW profile),
the \citet{Dalcanton10} correction produces pressure support that is very similar 
to the constant scale height ($\rho(R)\propto\Sigma(R)$, \citealt{Meurer96}, \citealt{Bouche22}) 
and \citealt{Weijmans08} cases (assuming constant dispersion), 
which all predict lower support corrections than for the self-gravitating disk case. 
This difference arises because these three cases assume constant scale height or a thin disk approximation, 
resulting in a correction of approximately $\D\ln\Sigma(R)/\D\ln{}R$. 
In contrast, the self-gravitating disk case explicitly assumes a constant vertical dispersion, 
so predicts $\rho(R)\propto\Sigma(R)^2$, yielding a correction term that is roughly twice that of the other cases.


\begin{figure}
\centering
\hspace{-7pt}
\includegraphics[width=0.49\textwidth]{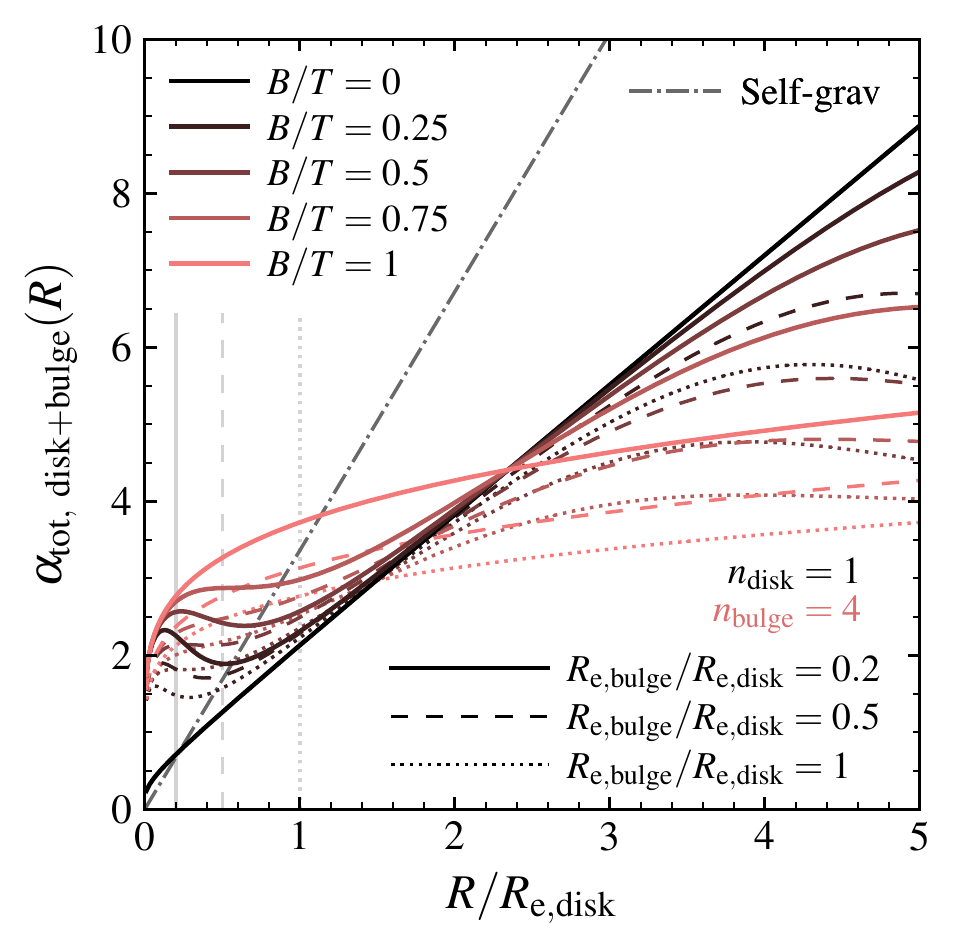} 
\caption{Composite pressure support correction, $\alphatot(R)$, for gas distributed in a composite disk+bulge system 
(with $\nSdisk=1$, $\nSbulge=4$), for a range of \bt (colors) and $\Rebulge/\Redisk$ ratios (dash length). 
For the limiting cases, we recover the profiles shown in Figure~\ref{fig:9}: 
$\bt=0$ has $\alphatot=\alpha(n=1)$ (black solid line), while $\bt=1$ has $\alphatot=\alpha(n=4)$ but with 
different radial scaling, owing to the different adopted $\Rebulge/\Redisk$ ratios (colored lines). 
For the cases with $0<\bt<1$, the bulge contribution modifies the $\alpha(n=1)$ profile at both small and large radii, 
leading to larger $\alphatot$ in inner regions and smaller $\alphatot$ in the outskirts ($R/\Redisk\lesssim,\gtrsim1-2$). 
At fixed \bt, the deviation from the disk $\alpha(n=1)$ in the center ($R/\Redisk\lesssim1-2$) is larger for smaller $\Rebulge/\Redisk$, 
while at large radii the deviation is larger for larger $\Rebulge/\Redisk$. 
For reference, we mark $\Rebulge/\Redisk$ with vertical light grey lines, and also show $\alphaSG$ (grey dash-dot line).}
\label{fig:11}
\end{figure}

These differences between $\alpha$ predict different pressure support-corrected $\vrot(R)$ for the same 
circular velocity profile and intrinsic velocity dispersion. We demonstrate these differences for $\alpha$ 
for deprojected S\'ersic models and a self-gravitating disk in Figure~\ref{fig:10}, 
over a range of S\'ersic indices ($n=0.5,1,2,4$; left to right) 
and intrinsic axis ratios ($\qint=1, 0.2$; top and bottom, respectively).\footnote{Though \alphan does not depend on 
the intrinsic axis ratio \qint, we show the velocity profiles for both a spherical and a flattened deprojected model
as an example of the composite effects from the variations to \vcirc and the pressure support distribution.} 
For all cases, we determine the circular velocity \vcirc (solid black line) assuming the mass distribution 
follows a single deprojected S\'ersic model of $\Mtot=10^{10.5}\Msun$ (i.e., a pure gas disk, or gas+stars where 
both components follow the same density distribution). 
We then calculate \vrot using both \alphan and \alphaSG (dashed and dotted lines, respectively). 
As implied by Figure~\ref{fig:9}, for $n\gtrsim1$ the rotation curves \vrot computed with \alphan 
are higher than with \alphaSG at $R\gtrsim\Reff$ (i.e., smaller correction from \vcirc). 
The difference between the two \vrot curves becomes more pronounced towards larger radii, 
in line with the continued decrease of $\alphan/\alphaSG$ with increasing radius. 
We also see the opposite behavior in the $n=0.5$ case, where \vrot 
computed in the self-gravitating case is higher than for \alphan at $R\gtrsim2.4\Reff$ 
(but the \vrot computed with \alphan is higher than with \alphaSG at smaller radii).

The amplitude of the intrinsic dispersion further impacts the \vrot profiles by causing disk truncation for 
sufficiently high \sigmaint relative to \vcirc, as previously discussed by \citet{Burkert16}. 
For the highest dispersion case ($\sigmaint=90\unit{km\,s^{-1}}$; orange), the pressure support correction 
predicts disk truncation (i.e., \hbox{$\vrot^2 \leq 0$}) within $R\lesssim5\Reff$ for both \alphan and \alphaSG. 
With medium dispersion ($\sigmaint=60\unit{km\,s^{-1}}$; turquoise), we still find disk truncation 
at $R\lesssim5\Reff$ for all $n$ when using \alphaSG, but only $\alpha(n=0.5,1)$ produce truncation within this radial range. 
Finally, \alphaSG does not produce truncation within $5\Reff$ in any case at the lowest dispersion 
($\sigmaint=30\unit{km\,s^{-1}}$; purple), and only $\alpha(n=0.5)$ predicts truncation at 
\hbox{$R\sim5\Reff$} (for both the spherical and flattened cases).

\subsection{Pressure support for multi-component systems}
\label{sec:asymmdrift_multi}

However, the gas in galaxies may be distributed in more than one component, 
which would modify the pressure support correction term.\footnote{Only the gas density 
distribution that impacts $\alpha(R)$, regardless of other (e.g. stellar or halo) components 
(see Section~\ref{sec:asymmdrift_single}).}
We can then derive the composite $\alphatot(R)$ using the $\alpha(R,n)$ of the individual gas components. 
For example, if the composite system includes gas in both a bulge and a disk, 
we have the total $\rho_{\mathrm{tot}}=\rho_{\mathrm{disk}}+\rho_{\mathrm{bulge}}$. 
\hbox{As $\D\ln\rho/\D\ln R = (R/\rho)(\D\rho/\D R)$,} we can write 
\begin{gather}
\frac{\D \ln \rho_{\mathrm{tot}}}{\D \ln R } = \frac{1}{\rho_{\mathrm{tot}}} 
\left( \rho_{\mathrm{disk}} \frac{\D \ln \rho_{\mathrm{disk}}}{\D \ln R } 
+ \rho_{\mathrm{bulge}} \frac{\D \ln \rho_{\mathrm{bulge}}}{\D \ln R}  \right), \ \mathrm{or} \nonumber \\
\alphatot(R) = \frac{1}{\rho_{\mathrm{tot}}} \left(\rho_{\mathrm{disk}}\alpha_{\mathrm{disk}} + 
\rho_{\mathrm{bulge}}\alpha_{\mathrm{bulge}}\right)
\label{eq:AD_composite}
\end{gather}
As discussed in Section~\ref{sec:asymmdrift_single}, this composite gas pressure support term 
is applicable to the gas velocity curve regardless of the distribution of the other, non-gas mass components.

We demonstrate an example composite pressure support term for a galaxy with gas distributed in both 
a disk and bulge over a range of \bt and $\Rebulge/\Redisk$ values in Figure~\ref{fig:11}. 
Here we assume $\nSdisk=1$ and $\nSbulge=4$ (i.e., exponential disk and de Vaucouleurs spheroid bulge, 
as adopted for recent bulge/disk decompositions at $z\sim1-3$ given the current 
observation spatial resolution limitations; \citealt{Bruce12}, \citealt{Lang14}), 
with a range of \bt (from disk- to bulge-only; black to light red colors) 
and $\Rebulge/\Redisk$ (from $\Redisk=5\Rebulge$ down to $\Redisk=\Rebulge$; solid to dotted line styles). 
As expected, the composite \alphatot is lower than \alphaSG at large radii, but can be larger than 
\alphaSG at $R/\Redisk\lesssim1$ when there is non-zero bulge contribution (see Figure~\ref{fig:9}).

Compared to the disk-only $\alpha(n=1)$ (solid black line), 
the inclusion of the bulge component leads to larger \alphatot at small radii ($R/\Redisk\lesssim1-2$) and 
lower \alphatot at large radii ($R/\Redisk\gtrsim1-2$). 
This is the result of a steeper inner density slope together with a shallower decline at large radii for $n=4$ 
compared to an exponential deprojected S\'ersic model,
so the bulge component becomes more important at very small and very large radii. 
When varying $\Rebulge/\Redisk$, we find the most pronounced changes to 
\alphatot at small radii when the $\Rebulge/\Redisk$ ratio is smallest (solid lines). 
This effect is less pronounced for larger $\Rebulge/\Redisk$ values, 
as the bulge density profile is more extended and the disk profile becomes important at smaller $R/\Redisk$. 
For larger radii, the opposite holds: the largest changes with $\bt$ are found for the largest 
$\Rebulge/\Redisk$ (dotted lines), as the bulge component becomes important at 
smaller $R/\Redisk$ owing to the larger $\Rebulge$.


\vspace{6pt}
\section{Discussion and implications}
\label{sec:disc}

In this paper we have presented properties and implications when using deprojected, axisymmetric S\'ersic models 
to describe mass density distributions or kinematics, over a wide range of possible galaxy parameters. 
Some of these effects will be more important for certain galaxy populations and epochs than others 
(as initially hinted in Figure~\ref{fig:8}). 
Here we discuss the implications for the models presented in this work, 
focusing on which aspects are most important for interpreting observations and for comparing observations to 
simulations as a function of cosmic time and galaxy mass.

\subsection{Low redshift}
\label{sec:disc_lowz}

Nearby, present-day star-forming galaxies (that are not dwarf galaxies) 
typically host fairly thin disk components, and where some also host a bulge. 
The disks of such galaxies would generally be characterized by geometries with small \qint --- 
relatively similar to the infinitely thin exponential disk case (\citealt{Freeman70}). 
Thus, when modeling the circular velocity curves of these disks, the choice of adopting 
the infinitely thin disk versus deprojected oblate S\'ersic models has a relatively small impact.
The thin gas disks of these local galaxies also 
have relatively low intrinsic velocity dispersions, with relatively little pressure support. 
The exact pressure support correction formulation therefore has less of an impact on the interpretation of the dynamics.

However, the low \qint and typically large disk effective radii \Redisk in $z\sim0$ star-forming galaxies, 
when coupled with a non-negligible bulge component, do result in ratios of \rhalftDbar/\Redisk less than 1. 
This deviation of the 2D and 3D half-mass radii can lead to large aperture effects when 
interpreting projected versus 3D quantities, such as when comparing observational or simulation quantities (e.g., \fDM). 
This aperture mismatch would be most severe for higher mass low-$z$ galaxies, as these will tend to 
have larger values of \Redisk and \bt (since a more prominent bulge will decrease \rhalftDbar relative to \Redisk). 
For example, aperture differences can lead to discrepancies of up to 
$\Delta\fDM = \fDM(\Redisk)-\fDM(\rhalftDbar) \sim0.15$ at $\Mstar\sim10^{11}\Msun$ 
for typical values of \Redisk and \bt (Figure~\ref{fig:8}, lower right).
In contrast, lower mass low-$z$ galaxies will generally be less impacted by aperture mismatches, 
owing to the lower typical \bt and smaller \Redisk of these galaxies.

Compared to the impact of aperture mismatches, 
definition differences in \fDM (as might be measured from observations and simulations) lead to only minor discrepancies. 
However, for lower stellar mass low-$z$ galaxies where the aperture mismatch is relatively minor, 
the typically low \bt and thus more prominent thin disk leads to a larger relative impact of \fDM estimator differences, 
as these galaxies are overall less spherically symmetric (see Figures~\ref{fig:6} \& \ref{fig:8}).

\vspace{6pt}

Overall, for star-forming galaxies at low redshift, the most important effect to consider is 
to correct for --- or avoid --- any aperture mismatches when comparing measurements between 
simulations and observations of, e.g., \fDM, particularly for high stellar masses. 
The impact of other aspects (use of infinitely thin disks vs. finite thickness, 
pressure support correction formulation, \fDM estimator definition) are all relatively minor 
and can be ignored for most purposes.

\subsection{High redshift}
\label{sec:disc_highz}

In contrast to the local universe, at 
high redshift (e.g., $z\sim1-3$) relatively massive star-forming galaxies 
generally exhibit thick disks, with increasing bulge contributions towards higher masses. 
These thick disks would be reasonably well described by elevated $\qint\sim0.2-0.25$. 
As the derived circular velocity curve for such a geometry is fairly different from that of 
an infinitely thin exponential disk (e.g., \citetalias{Noordermeer08}), the choice of rotation curve parameterization 
(i.e., adopting \vcirc based on a deprojected profile such as those presented here versus using 
an infinitely thin exponential disk) is important at high-$z$.

The thick geometries of high-$z$ disks are coupled with relatively high intrinsic velocity dispersions, 
which implies that the overall amount of pressure support 
is expected to be much higher than for the dynamically-cold, thin disks at low-$z$. 
Thus, not only is accounting for pressure support more important, 
but the choice of adopted pressure support correction matters much more 
for interpreting kinematics at high-$z$ than for nearby galaxies.

In this paper, we have derived the log density slope-driven pressure support correction \alphan 
as a function of radius R for the deprojected S\'ersic models, and have compared this correction term to other formulations, 
particularly the correction for a self-gravitating exponential disk, \alphaSG (as in \citealt{Burkert10,Burkert16}).
A key implication of the differences between these pressure support corrections is that, 
for the same $\vrot(R)$ and \sigmaint, \alphan predicts a lower \vcirc than would be inferred when applying \alphaSG 
(i.e., the inverse of the demonstration in Fig.~\ref{fig:10}). 
Furthermore, the shape of the inferred \vcirc profile can also differ 
(particularly when considering a composite disk+bulge gas distribution; Fig.~\ref{fig:11}). 
Both effects can impact the results of mass decomposition from modeling of galaxy kinematics, 
which have important implications for the measurement of dark matter fractions.

Though the smaller disk sizes of high-$z$ galaxies help to alleviate the disk-halo degeneracy that strongly impacts 
kinematic fitting at $z\sim0$, there are nonetheless often degeneracies between mass components 
when performing kinematic modeling at $z\sim1-3$ (see e.g., \citealt{Price21}, Sec.~6.2 \& Fig.~5). 
The strong pressure correction from \alphaSG can further complicate 
the reduced but still present disk-halo degeneracy at high-$z$. 
When combined with high \sigmaint, modest variations in \sigmaint (allowed within the uncertainties) 
can extend the degeneracy between galaxy-scale dark matter fractions and total baryonic masses 
--- in the most extreme cases, allowing the $1\sigma$ region to extend from 0\% to 50+\% dark matter fractions.

However, the strength of this added degeneracy effect depends not only on \sigmaint, 
but also on the pressure support prescription. The large correction from \alphaSG can result in a falling 
\vrot even for a flat or rising \vcirc (with a large halo contribution; Fig.~5b of \citealt{Price21}). 
Alternatively, if \alphan were adopted, the comparable correction 
to \vcirc would produce a less steeply dropping (or potentially flat) \vrot profile. 
Thus, to match the \emph{observed} \vrot profile, the intrinsic \vcirc would be limited to lower amplitudes 
(i.e., implying lower dynamical masses) 
with less shape modification than when using \alphan instead of \alphaSG. 
This in turn implies partial breaking of the added pressure support impact to 
the disk-halo degeneracy, restricting the higher likelihood regions towards 
\emph{\underline{lower}} \fDM. 
While adopting \alphan would have the greatest impact on the objects with high \sigmaint 
(where the pressure support has the largest impact), 
the change in prescription should impact the inferred mass distribution for all objects to some extent. 
The choice of pressure support formulation is thus an important factor in the interpretation of dynamics 
of high-redshift galaxies, and has direct implications for the interpretation of mass fractions. 
Overall, this will have the largest impact for galaxies with low $\vrot/\sigmaint$ 
(and the smallest impact for high $\vrot/\sigmaint$) as this will lead to the largest fractional change in \vrot relative to \vcirc.
Since there is currently no observed trend of \sigmaint with \Mstar at high-$z$ (e.g., \citealt{Ubler19}; 
though the dynamic range of \Mstar is currently limited), the correlation of \vrot with \Mstar would then 
cause this effect to generally be most important for low-mass galaxies.

On the other hand, the higher \qint and lower \Redisk of high-$z$ disk galaxies 
implies that aperture effects arising from deviations of \rhalftDbar versus \Redisk are 
less important than for low-$z$ galaxies, as the disk and bulge sizes are more similar. 
Still, there can be up to $\sim20\%$ radii aperture differences in the 2D and 3D half-mass radii 
(though only $\sim2.5\%$ \fDM differences), so depending on 
the particular measurement quantity and accuracy required, this effect could still be important. 
As is the case for the local limit, the aperture radii difference (and the resulting impact on inferred \fDM) 
typically has a larger impact for higher mass objects, 
since these tend to have higher \bt and \Redisk than lower mass objects. 
Finally, as with the low-$z$ case, the \fDM estimator differences are relatively minor compared to the 
other effects and can be generally ignored (though the same comments on trends with \bt and necessary 
comparison accuracy from the low-$z$ discussion apply in this case).

\vspace{6pt} 

In conclusion, for high redshift star-forming galaxies, the most important effects to consider are 
\begin{enumerate}[label=\textbf{\arabic*.}, topsep=4pt, itemsep=3pt, left=8pt]
\item adopting circular velocity curves that account for the finite, thick-disk geometry, and 
\item including a reasonable pressure support correction when interpreting rotation curves. 
\end{enumerate} 
In this limit (higher \qint, lower \Redisk, high \sigmaint), the other aspects 
(2D vs. 3D half-mass radii apertures, \fDM estimator definitions) 
have relatively small impacts and can typically be ignored.


\section{Summary}
\label{sec:summary}

We have presented a number of properties for 3D deprojected S\'ersic models 
with a range of intrinsic axis ratio $\qint = c/a$ (i.e., flattened/oblate, spherical, or prolate). 
We follow the derivation of \citetalias{Noordermeer08}, who presented the 
deprojection of the 2D S\'ersic profile to a 3D density distribution $\rho(m)$, 
as well as the midplane circular rotation curve $\vcirc(R)$ for such a mass distribution. 
We then extend this work by numerically deriving spherical enclosed mass profiles $\Menc(<r=R)$ 
and the log density slope $\D\ln\rho/\D\ln{}R$.

Using these profiles, we determine a range of properties of these mass models. 
Specifically, we examine the differences between the 2D projected effective radius, \Reff, 
and the 3D spherically-enclosed half-mass radius, \rhalftD, over a range of intrinsic axis ratios \qint and S\'ersic indices $n$, 
and find $\rhalftD>\Reff$, with the ratio approaching unity as $\qint\to0$, in agreement with previous results.
We also calculate virial coefficients that relate the circular velocity to either the total mass (\ktot) 
or the enclosed mass within a sphere (\kthrd).

Furthermore, we calculate derived properties for example composite galaxy systems 
(consisting of both flattened deprojected S\'ersic and spherical components), to consider 
how varying galaxy properties (i.e., \bt, \Redisk, $z$) impacts these properties, such as $\rhalftDbar/\Redisk$. 
We also examine the impact of different methods of inferring $\fDM(<R)$ 
and the compounding effects from measuring $\fDM$ within different aperture radii. 
We find that using different apertures, such as \rhalftDbar versus \Redisk, 
can lead to very large differences in the measured \fDM, particularly for high \bt and low $\Rebulge/\Redisk$. 
In contrast, using different \fDM definitions, such as 
$\fDMv(<R)=\vcircDM^2(R)/\vcirctot^2(R)$ and $\fDMm(<R)=\MencDM(<r=R)/\Menctot(<r=R)$, 
only produces minor differences when measured at the same radius. 
Using toy models, we estimate how $\rhalftDbar/\Redisk$ and the \fDM estimators (measured both at \Redisk 
and with mismatched \rhalftDbar vs. \Redisk apertures) change as a function of redshift and stellar mass, 
and find increasing offsets towards higher \Mstar and lower $z$.

We additionally use the deprojected S\'ersic models to derive self-consistent pressure support correction terms, 
with $\alpha(R,n) = -\D\ln\rho_g(R,n)/\D\ln{}R$ for constant gas velocity dispersion. 
We find that at $R\gtrsim\Reff$, $\alpha(R,n)$ typically predict a smaller pressure support 
correction than is inferred for a self-gravitating disk (as in \citealt{Burkert10,Burkert16}), 
and are more similar to predictions derived for thin disks with $\sim$constant scale heights under various assumptions 
(e.g., \citealt{Dalcanton10}, \citealt{Meurer96}, \citealt{Bouche22}, \citealt{Weijmans08}) 
and from simulations (e.g., \citealt{Kretschmer21}; also \citealt{Wellons20}). 
The effect of a lower pressure support with \alphan implies larger \vrot for the same \vcirc and \sigmaint{} 
(or lower \vcirc for the same \vrot and \sigmaint)than if assuming \alphaSG, and would predict any disk truncation 
(where $\vrot\to0$, as in \citealt{Burkert16}) at larger radii than for the self-gravitating case.

Finally, we discuss implications of this work for future studies of galaxy mass distributions and kinematics. 
Low-$z$ star-forming disk galaxies typically have thin disks with small \qint and low intrinsic velocity dispersion, 
so the most important effect to consider is aperture mismatches when comparing measurements --- 
such as measuring \fDM within 2D and 3D apertures, as typically adopted for observations and simulations, respectively. 
In contrast, the thick disks in high-$z$ star-forming galaxies are characterized by large \qint and high intrinsic 
velocity dispersion, so adopting circular velocity curves accounting for this finite thickness 
and accounting for the pressure support correction are the most important aspects. 
The large \sigmaint of these high-$z$ galaxies can produce large pressure support corrections, 
in some cases causing greater-than-Keplerian falloff in outer rotation curves (e.g., \citealt{Genzel17}). 
In this limit of relatively large correction amplitudes, the choice of the adopted pressure support correction is also important 
and can impact constraints of the disk-halo mass decomposition, 
as lower correction amplitudes (e.g., using \alphan versus the larger correction of \alphaSG) 
will tend to lead to lower inferred dark matter fractions, particularly for high \sigmaint. 
Furthermore, while differences in quantity estimators (e.g., \fDMm vs. \fDMv) have only modest effects at both 
low and high-$z$, as measurements improve it would be worth correcting for, or avoiding, estimator 
differences to improve the accuracy of comparisons between different studies.

The deprojected S\'ersic profile models presented here 
can be used to aid comparisons between observations and simulations, 
to help convert between simulation quantities that are typically determined within spherical shells 
and observational constraints based on 2D projected quantities. 
As demonstrated in this work, commonly adopted apertures for simulations (3D half-mass) 
versus observations (2D projected half-light or half-mass) can probe different physical scales, 
impacting observation-simulation comparisons, particularly for dark matter fractions. 
The pre-computed profiles and values (or similar calculations) can help to move towards more direct, 
apples-to-apples comparisons between the two, without resorting to the more direct but complex step of 
constructing and analyzing mock observations based on simulated galaxies 
(as in, e.g., \citealt{Ubler21}; but see also \citealt{Genel12a}, \citealt{Teklu18}, \citealt{Simons19}).
The code used to compute these profiles, as well as precomputed profiles and other quantities 
for a range of S\'ersic index $n$ and intrinsic axis ratio \qint, have been made publicly available.


\begin{acknowledgements}

We thank Michael Kretschmer for sharing the values of $\alpha_{\rho}$ derived from the simulated galaxies 
presented in Fig.~4 of \citet{Kretschmer21}, Taro Shimizu for helpful discussions, 
and Dieter Lutz for comments on the manuscript. 
We also thank the anonymous referee for their comments and suggestions that improved this manuscript. 

H{\"U} gratefully acknowledges support by the Isaac Newton Trust and 
by the Kavli Foundation through a Newton-Kavli Junior Fellowship.

This work has made use of the following software: 
Astropy\footnote{http://www.astropy.org}  \citep{astropy:2013, astropy:2018}, 
dill \citep{McKerns11, pathosMcKerns}, 
IPython \citep{Perez07}, 
Matplotlib \citep{Hunter07}, 
Numpy \citep{van2011numpy, 2020NumPy}, 
Scipy \citep{2020SciPy}

\end{acknowledgements}

\end{document}